\documentclass{IEEEtran}
\usepackage[utf8]{inputenc}
\usepackage{rotating}
\usepackage{booktabs}
\usepackage[usenames, dvipsnames]{color}
\usepackage[T1]{fontenc} 
\usepackage{amsmath}
\usepackage{epsfig,endnotes}
\usepackage[colorinlistoftodos,prependcaption,textsize=tiny]{todonotes}
\usepackage{hyperref}

\usepackage{mathtools}
\usepackage{url}
\usepackage[noend]{algpseudocode}
\usepackage{tabularx,colortbl}
\usepackage{multirow}
\usepackage[Symbolsmallscale]{upgreek}
\usepackage{algorithm}
\usepackage{algpseudocode}
\usepackage{mathtools}
\usepackage{url}
\usepackage{authblk}
\usepackage{tablefootnote}
\usepackage{siunitx}
\usepackage{listings}
\usepackage{amssymb}
\usepackage{bbding}

\title{SoK - Security and Privacy in the Age of Drones: Threats, Challenges, Solution Mechanisms, and Scientific Gaps}
\author{
Ben Nassi$^{1}$, Asaf Shabtai$^{1}$, Ryusuke Masuoka$^{2}$, Yuval Elovici$^{1}$\\ 
nassib@post.bgu.ac.il, shabtaia@post.bgu.ac.il, masuoka.ryusuke@jp.fujitsu.com, elovici@inter.net.il \\
$^{1}$Ben-Gurion University of the Negev, $^{2}$Fujitsu Systems Integration Laboratories
}

\begin{document}
\maketitle

\section*{Abstract}
The evolution of drone technology in the past nine years since the first commercial drone was introduced at CES 2010 has caused many individuals and businesses to adopt drones for various purposes. We are currently living in an era in which drones are being used for pizza delivery, the shipment of goods, and filming, and they are likely to provide an alternative for transportation in the near future. However, drones also pose a significant challenge in terms of security and privacy within society (for both individuals and organizations), and many drone related incidents are reported on a daily basis. These incidents have called attention to the need to detect and disable drones used for malicious purposes and opened up a new area of research and development for academia and industry, with a market that is expected to reach \$1.85 billion by 2024. While some of the knowledge used to detect UAVs has been adopted for drone detection, new methods have been suggested by industry and academia alike to deal with the challenges associated with detecting the very small and fast flying objects. 
In this paper, we describe new societal threats to security and privacy created by drones, and present academic and industrial methods used to detect and disable drones. We review methods targeted at areas that restrict drone flights and analyze their effectiveness with regard to various factors (e.g., weather, birds, ambient light, etc.). We present the challenges arising in areas that allow drone flights, introduce the methods that exist for dealing with these challenges, and discuss the scientific gaps that exist in this area. Finally, we review methods used to disable drones, analyze their effectiveness, and present their expected results. Finally, we suggest future research directions and discuss whether the benefits from the decision to allow drones to fly in populated areas are actually worth the risk.

\section{Introduction}
"Terror by Joystick" \cite{terrorism-by-joystick-1,terrorism-by-joystick-2,terrorism-by-joystick-3}and "Eyes in the Skies " are examples of topics  that have been provided by the media as a means of describing the malicious impact of drones today. There is no doubt that drones are a game-changing technology in terms of security and privacy, and have become a growing societal threat in recent years. Just a decade ago, drones were considered a technology restricted for use by official authorities such as the military, police, etc. However, in the nine years since the first commercial drone was presented at CES 2010 by Parrot, many sectors have begun to use drones (including the private sector), and drone shipments are expected to reach 805K by 2021 due to their reasonable price and diverse uses.

In addition to their increased adoption by the industrial and private sectors \cite{PIZZA-DELIVERY-BY-DRONE-LAUNCHED-BY-DOMINOS, Amazon, CNN-Just-Got-Approved-to-Fly-Drones-Over-Crowds-of-People}, drones have also been adopted by many entities for various malicious purposes, and drone related incidents are reported on a daily base \cite{cheating, 26-floor, Woman-grabs-gun-shoots-nosy-neighbour-s-drone, Virginia-Woman-Shoots-Down-Drone-Near-Actor-Robert-Duvalls-Home, not-my-backyard-man-arrested-after-shooting-drone-down, Big-rise-in-drone-jail-smuggling-incidents, Two-plead-guilty-in-border-drug-smuggling-by-drone, smuggling-iphones, smuggling-tobaco, venezuela}. The volume of drone related incidents will likely increase further along with the expected growth in drone shipments in the coming years \cite{drone-marketshare,drone-marketshare-2} and the new "open sky" policy adopted by many countries (US \cite{President-Trump-Moves-to-Fill-America's-Skies-with-Drones}, UK \cite{Amazon}, New Zealand \cite{PIZZA-DELIVERY-BY-DRONE-LAUNCHED-BY-DOMINOS}) which allows drones to fly over populated areas.

The growing number of incidents has highlighted the need to detect and disable drones that are maliciously used by their operators and has created a new avenue of drone research and development for academia and industry focused on  anti-drone methods. The anti-drone market is expected to reach \$1.85 billion by 2024 \cite{anti-drone-market}, and many solutions have already been suggested by researchers and companies to: (1) detect nearby drones and issue alerts about their presence,  and (2) disable them. While some of the knowledge used to develop these solutions was adopted from the related area of UAV detection, many other methods have been developed specifically for drones due to the challenges that arise from their small size and versatility which make detecting and disabling them more difficult than detecting and disabling a UAV.

In this paper, we discuss security and privacy in the era of drones. First, we describe new threats that drones pose to society and future threats that are on the horizon due to recent technological improvements. We review methods to detect drones in areas that restrict drone flights and analyze their effectiveness with regard to various factors (e.g., weather, birds, ambient light, etc.). We continue by describing two new challenges that have arisen in areas that allow drone flights, and review existing methods for dealing with these challenges and discuss the scientific gaps that exist in this area. In addition, we review methods to disable drones, analyze their effectiveness, and present their expected results . At the end of this paper, we suggest future research directions that should be investigated in order to improve societys ability to handle the threats posed by drones.

\textbf{Contribution} While a SoK in this area has already taken place \cite{altawy2017security, guvenc2018detection, guvencc2017detection, game-of-drones, sturdivant2017systems, shi2018anti, michel2018counter}, we consider previous attempts very limited compared to our SoK, because (1) they either ignore industry methods  \cite{altawy2017security, guvenc2018detection, guvencc2017detection,game-of-drones, sturdivant2017systems, shi2018anti} or academic methods \cite{game-of-drones,michel2018counter}, (2) they don't review methods to deal with the challenges that arise from allowing drones to fly over populated areas \cite{altawy2017security, guvenc2018detection, guvencc2017detection, game-of-drones, sturdivant2017systems, game-of-drones, shi2018anti}, and (3)they are limited in scope, failing to review the large number of methods covered in this paper or analyze their effectiveness. In our SoK, we review 120 methods proposed by the academic and industrial sectors that were designed to detect and disable drones flying in areas where drone presence is restricted, as well as areas where drones are allowed. We compare the methods' effectiveness at drone detection. We also present the scientific gaps that exist as a result of allowing drones to fly over populated areas and discuss future research directions.

\textbf{Structure} The remainder of the paper is structured as follows: in section \ref{Sec:background}, we describe drone types, architecture, and functionality. In section \ref{Sec:Threat Models}, we discuss the current threats posed by drones. In sections \ref{Sec:detection-restricted} and \ref{sec:detection-non-restricted}, we review methods that have been proposed by the academic and industrial sectors to detect drones in restricted areas (e.g., critical infrastructure, airports) and in non-restricted areas (e.g., countries that apply an "open skies" policy). Methods to disable drones and countermeasures are reviewed in section  \ref{sec:Disabling}, and in section \ref{sec:future-research-directions}, we discuss areas in which there is a scientific gap and propose future research directions.

\section{Drones - Background}
\label{Sec:background}
In this section, we provide the relevant background required for the rest of the paper. Drones are multirotor aircraft operated by a controller. Drones are sometimes named according to the number of rotors  they have: tricopter, quadcopter, hexacopter, and octocopter are frequently used to refer to three, four, six, and eight rotor rotorcraft, respectively. In this paper, we refer to all of them as drones. Until 2010, drones were built by amateurs and used primarily by hobbyists for fun. In the nine years since the first commercial drone was presented by Parrot at CES 2010, drones have been adopted by industry for various purposes \cite{PIZZA-DELIVERY-BY-DRONE-LAUNCHED-BY-DOMINOS, Amazon, CNN-Just-Got-Approved-to-Fly-Drones-Over-Crowds-of-People} and are considered by many as a game-changing technology that is here to stay. In the subsections that follow, we describe drone types, architecture (communication, sensors, and functionalities), and use. Understanding drone technology and capabilities is important for appreciating the challenges they create and developing methods for detecting and disabling them. 
In  the next subsection \ref{subsection:categories}, we introduce the categories of commercial drones. In this paper, we focus on three categories: Nano, Micro, and Mini Drones. In the subsequent subsections \ref{subsection:fpv}, \ref{subsection:sensors}, \ref{subsection:functionality} we provide necessary information about their architecture; in subsection \ref{subsection:use}, we discuss the ways each type of drone is currently used.

\begin{figure*}[]
  \centering
  \begin{minipage}[b]{0.34\textwidth}
\includegraphics[width=\textwidth]{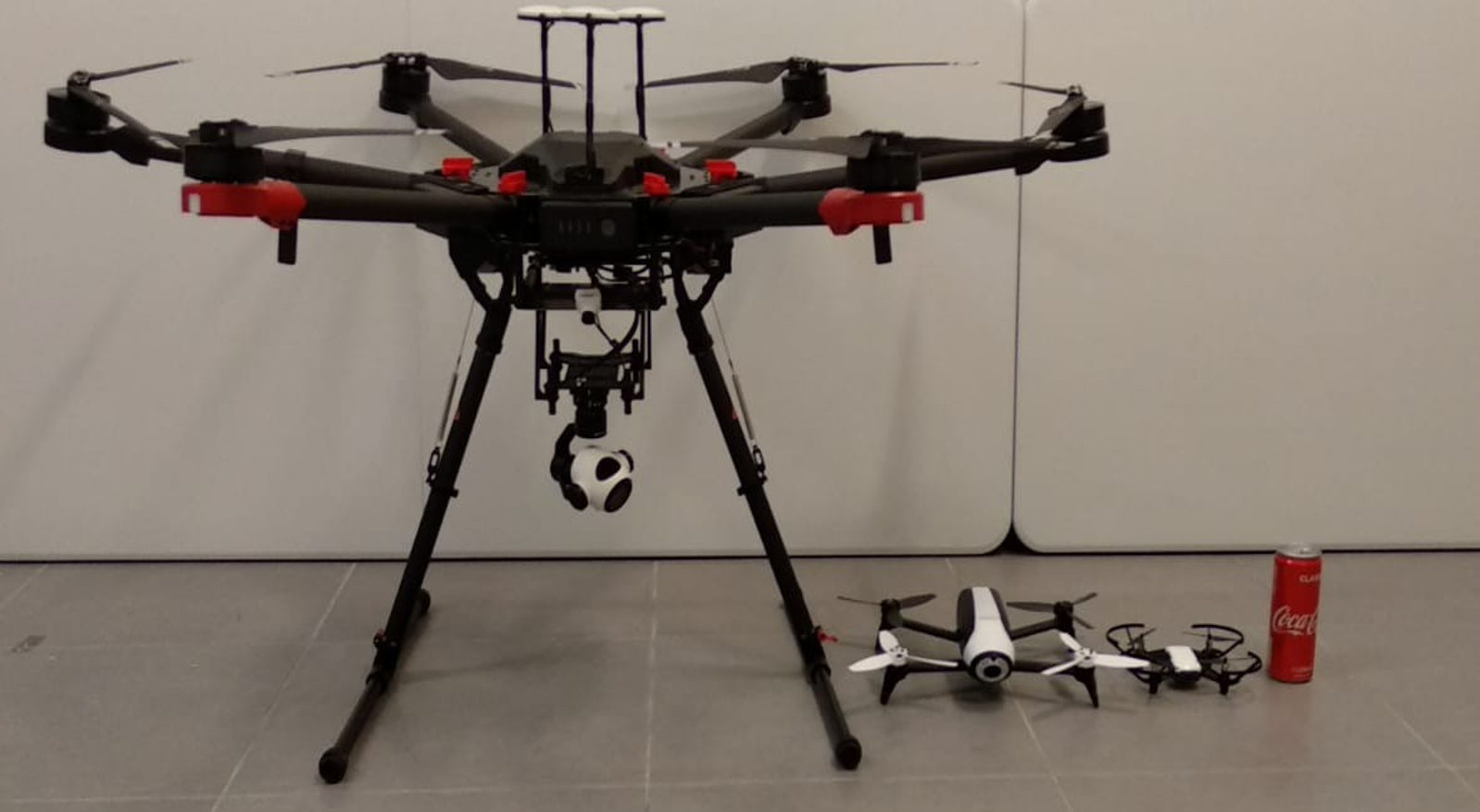}
\caption{Mini (DJI Matrice 600 Pro), Micro (Parrot Bebop 2) and Nano (DJI Tello) drones compared to the size of a Coca-Cola can.}
\label{fig:sizes}
  \end{minipage}
  \hspace{0.05cm}
  \begin{minipage}[b]{0.64\textwidth}
    \includegraphics[width=\textwidth]{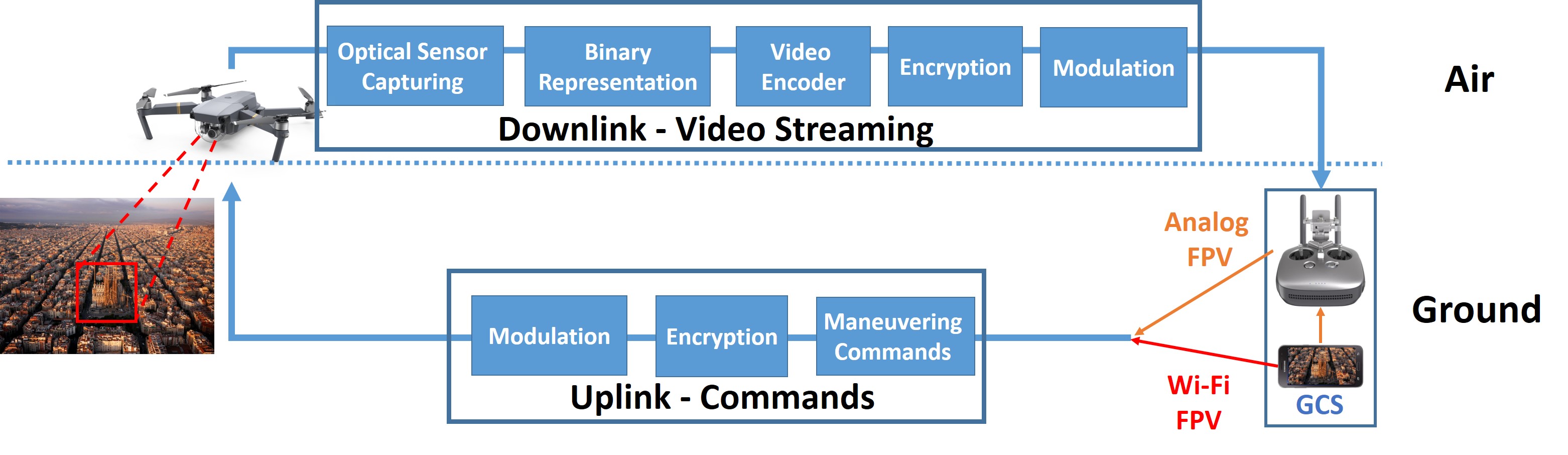}    
    \caption{FPV channel - downlink and uplink.}
    \label{fig:FPV}
  \end{minipage}
\end{figure*}


\subsection{Types \& Categories}
\label{subsection:categories}
Various organizations (NATO, DoD , NASA, State  Regulatory Authority) have defined groups or classes of drones. Most classification has been made based on weight, altitude, or speed. While classification differs among these organizations, the most common classification system used (based on drone weight) defines five groups of drones: Nano, Micro, Mini, Small, and Tactical. Table \ref{tab:categories} presents the weight, payload, and range for each of these groups. Nano drones (e.g., Eachine E10C, Holy Stone HS210) are smaller than a human hand. Macro (e.g., DJI Spark, DJI Mavic, Parrot Bebop 2) and Mini (e.g., DJI Matrice 600 Pro, DJI Inspire 2) are larger, as can be seen in Figure \ref{fig:sizes} which presents Mini (DJI Matrice 600 Pro), Micro (Parrot Bebop 2), and Nano (DJITello) drones next to a can of Coke. Small and Tactical drones are the size of a motorcycle/helicopter and are used for military purposes and will likely provide an alternative for transportation (manual \cite{hoversurf} and automatic \cite{SureFly,ehang})  in the future. In this paper, we focus on Nano, Micro, and Mini drones that are mainly used for civilian purposes.

\begin{table}[]
\caption{Types Of Drones}
\label{tab:categories}
\resizebox{1.0\linewidth}{!}{%
\begin{tabular}{|l|l|l|l|l|}
\hline
\textbf{Category} & \textbf{Weight} & \textbf{\begin{tabular}[c]{@{}l@{}}Operating\\ Altitude\end{tabular}} & \textbf{Range} & \textbf{Payload} \\ \hline
Nano & \textless 0.2 kg & \textless 90 & 90 m & \textless 0.2 kg \\ \hline
Micro & 0.25-2 kg & \textless 90 m & 5 km & 0.2-0.5 kg \\ \hline
Mini & 2-20 kg & \textless 900 m & 25 km & 0.5-10 kg \\ \hline
Small & \textless 150 kg & \textless 1500 m & 50-100 km & 5-50 kg \\ \hline
Tactical & \textgreater 150 kg & \textless 3000 m & \textgreater 200 km & 25-200 kg \\ \hline
\end{tabular}
}
\end{table}


\subsection{First-Person View (FPV) Channel}
\label{subsection:fpv} 
Modern drones provide video piloting capabilities (an FPV channel) based on radio signals, in which a live video stream is sent from the drone's video camera to the pilot (operator) via a GCS (ground control station) on the ground (dedicated controller, smartphone, VR glasses, smartwatch). The FPV channel enables the pilot to fly the drone as if he/she was on board (instead of looking at the drone from the pilot's actual ground position). The FPV channel also allows an operator to control a drone using a GCS. A typical FPV channel consists of an uplink and a downlink, as demonstrated in Figure \ref{fig:FPV}. A video downlink is used for video streaming using data that is captured by the drone's camera and sent to the pilot's GCS screen. The video streaming process usually consists of digitization of the captured picture to binary representation by a CMOS sensor, which is followed by video compression, encryption, and modulation. Real-time protocols for video streaming are used for these purposes. The second channel, referred to as an uplink, is largely used for C\&C of the drone, and the uplink process usually consists of digitizing the joystick's movements (or smartphone) to binary commands, which is followed by encryption and modulation. By its nature,  the amount of data that is sent over a video downlink is much greater than that sent by a remote control downlink.

There are two types of technologies dominating the FPV market: Wi-Fi FPV and analog FPV, and two types of technologies dominating the FPV market: Wi-Fi FPV and analog FPV \cite{WiFi-FPV-vs-5.8GHz-FPV-vs-2.4GHz-FPV-Ultimate-Guide}. A \textbf{Wi-Fi FPV} drone is basically a flying router  with no Internet connectivity. Such drones open a flying router with no Internet connectivity. Such drones open a network (access point) that allows the drone and its controller to communicate. Wi-Fi is by far the most popular method used to include FPV in budget RC drones (according to \cite{WiFi-FPV-vs-5.8GHz-FPV-vs-2.4GHz-FPV-Ultimate-Guide, WiFi-FPV-vs-5.8GHz-FPV-vs-2.4GHz-FPV}), because: (1) any Android/iOS smartphone (or tablet) on the market can be used to operate the drone, (2) the only additional hardware required is a Wi-Fi FPV transmitter (which is connected to the camera of the drone), instead of an additional controller with a screen that is equipped with a dedicated radio transceiver which is required by other FPV types (e.g., 2.4/5.8 GHz analog FPV), (3) drone manufacturers can boost the Wi-Fi FPV drone flight range up to four kilometers (e.g., DJI Spark) using dedicated hardware, and (4) Wi-Fi FPV drones support 4K resolution. Some drone types are considered pure Wi-Fi FPV drones (e.g., DJI Spark, DJI Phantom 3 SE, Parrot Bebop 2), and other types of drones contain Wi-Fi FPV along with their dedicated analog FPV (e.g., DJI Mavic pro, DJI Mavic Air). Almost every FPV-enabled drone selling for less than \$100 uses Wi-Fi FPV \cite{WiFi-FPV-vs-5.8GHz-FPV-vs-2.4GHz-FPV-Ultimate-Guide}, and there are dozens of kinds of Wi-Fi FPV drones available for purchase \cite{Wifi-FPV-Drones,8-smartphone-controlled-drones, 8-DRONES-THAN-CAN-BE-CONTROLLED-BY-A-SMARTPHONE}, ranging from \$30 to hundreds and thousands of dollars. Table \ref{tab:osi} presents the OSI model of the DJI Spark. \textbf{Analog FPV} requires a dedicated GCS for mediation between a drone and a smartphone (that is used as a screen), and therefore drones with analog FPV are more expensive than those with Wi-Fi FPV. However, analog FPV drones can reach a distance of up to seven kilometers from their GCS (using dedicated amplifiers and remote controllers), supporting flight ranges of up to eight kilometers (DJI Matrice 600 Pro).

\begin{table}[]
\centering
\caption{OSI model of DJI Spark (Wi-Fi FPV drone)}
\label{tab:osi}
  \resizebox{0.55\linewidth}{!}{%
\begin{tabular}{|l|l|}
\hline
Layer 7 - Application & Linux \\ \hline
Layer 6 - Presentation & \multirow{2}{*}{RTP} \\ \cline{1-1}
Layer 5 - Session &  \\ \hline
Layer 4 - Transport & UDP \\ \hline
Layer 3 - IP & IPv4 \\ \hline
Layer 2 - Data Link & \multirow{2}{*}{\begin{tabular}[c]{@{}l@{}}IEEE\\ 802.11n\end{tabular}} \\ \cline{1-1}
Layer 1- Physical &  \\ \hline
\end{tabular}
}
\end{table}

\subsection{Sensors}
\label{subsection:sensors}

The following sensors can be found in most of the commercial drones sold by the three biggest commercial drone manufacturers (DJI, Parrot and Yuneec).

\begin{itemize}
    \item Motion Sensors – Accelerometers, gyroscopes, and magnetometers provide nine dimensions of freedom and are used by the drone for stabilization.
    \item 4K/Full HD Video Camera – Such cameras are used to capture video and still images.
    \item GPS Device – This device is used by the drone for localization and automatic navigation.
    \item Barometer – This measuring device is used by the drone to calculate  its altitude during flight.
    \item Collision Avoidance System – Recent DJI drones are equipped with vision systems based on monocular cameras and ultrasonic sensors in order to sense dangers in the vicinity of the drone. Skydio \cite{Skydio} uses computer vision algorithms that obtain data from 13 integrated cameras for this task. 
\end{itemize}

\begin{table*}[]
\caption{Comparison of top commercial drones }
\label{tab:cmp-drones}
  \resizebox{1.0\linewidth}{!}{%
\begin{tabular}{cc|c|c|c|c|c|c|c|c|c|c|c|c|c|c|}
\cline{3-16}
                                              &                                                            & \multicolumn{2}{c|}{FPV} & \multicolumn{4}{c|}{Functionality}                                                                                                                                                    & \multicolumn{8}{c|}{Characteristics}                                                                                                                                                                                                                                                                                                                                                    \\ \hline
\multicolumn{1}{|c|}{Manufacturer}            & Name                                                       & Wi-Fi      & Analog      & \begin{tabular}[c]{@{}c@{}}Follow \\ Me\end{tabular} & RTH & \begin{tabular}[c]{@{}c@{}}Automatic\\ Navigation\end{tabular} & \begin{tabular}[c]{@{}c@{}}Smart\\ Capture\end{tabular} & \begin{tabular}[c]{@{}c@{}}Flight \\ Range \\ (km)\end{tabular} & \begin{tabular}[c]{@{}c@{}}Speed \\ (km/h)\end{tabular} & \begin{tabular}[c]{@{}c@{}}Flight \\ Time \\ (m)\end{tabular} & \begin{tabular}[c]{@{}c@{}}Weight \\ (g)\end{tabular} & \begin{tabular}[c]{@{}c@{}}Altitude \\ (km)\end{tabular} & \begin{tabular}[c]{@{}c@{}}Video \\ Resolution\end{tabular} & Date & \begin{tabular}[c]{@{}c@{}}Price \\ (\$)\end{tabular} \\ \hline
\multicolumn{1}{|c|}{\multirow{6}{*}{DJI}}    & \begin{tabular}[c]{@{}c@{}}Matrice \\ 600 Pro\end{tabular} &             & \Checkmark           & \Checkmark                                                    & \Checkmark   & \Checkmark                                                              &                                                         & 5                                                               & 65                                                      & 32 -38                                                        & 9500                                                  & 2.5-4.5                                                  & 4K         & 2016 & 4999                                                 \\ \cline{2-16} 
\multicolumn{1}{|c|}{}                        & \begin{tabular}[c]{@{}c@{}}Mavic 2 \\ Pro\end{tabular}     & \Checkmark          & \Checkmark           & \Checkmark                                                    & \Checkmark   & \Checkmark                                                              & \Checkmark                                                       & 8                                                               & 72                                                      & 31                                                            & 907                                                   & 6                                                        & 4K         & 2018 & 1499                                                 \\ \cline{2-16} 
\multicolumn{1}{|c|}{}                        & Spark                                                      &\Checkmark         &             &\Checkmark                                                   &\Checkmark  &\Checkmark                                                             &\Checkmark                                                      & 2                                                               & 50                                                      & 16                                                            & 300                                                   & 4                                                        & 1080p      & 2017 & 399                                                  \\ \cline{2-16} 
\multicolumn{1}{|c|}{}                        & Inspire 2                                                  &            &\Checkmark          &\Checkmark                                                   &\Checkmark  &\Checkmark                                                             &                                                         & 7                                                               & 94                                                      & 23-27                                                         & 3440                                                  & 2.5-5                                                    & 4K         & 2016 & 2999                                                 \\ \cline{2-16} 
\multicolumn{1}{|c|}{}                        & \begin{tabular}[c]{@{}c@{}}Phantom \\ 4 Pro\end{tabular}   &            &\Checkmark          &\Checkmark                                                   &\Checkmark  &\Checkmark                                                             &\Checkmark                                                      & 7                                                               & 72                                                      & 30                                                            & 1388                                                  &                                                          & 4K         & 2016 & 1499                                                 \\ \cline{2-16} 
\multicolumn{1}{|c|}{}                        & \begin{tabular}[c]{@{}c@{}}Mavic \\ Air\end{tabular}                                                  &\Checkmark         &\Checkmark          &\Checkmark                                                   &\Checkmark  &\Checkmark                                                             &\Checkmark                                                      & 2                                                               & 68.4                                                    & 21                                                            & 430                                                   & 5                                                        & 4K         & 2018 & 799                                                   \\ \cline{2-16} 
\multicolumn{1}{|c|}{}                        & \begin{tabular}[c]{@{}c@{}}Tello\end{tabular}                                                  &\Checkmark         &          &                                                &  &                                                            &                                                      & 0.1                                                               & 28.8                                                    & 13                                                            & 80                                                   & 0.1                                                        & 720p         & 2018 & 99                                                  \\ \hline
\multicolumn{1}{|c|}{\multirow{2}{*}{Parrot}} & Bebop 2                                                    &\Checkmark         &             &\Checkmark                                                   &\Checkmark  &\Checkmark                                                             &                                                         & 2                                                               & 64                                                      & 30                                                            & 525                                                   & 0.1                                                      & 1080p      & 2017 & 399                                                  \\ \cline{2-16} 
\multicolumn{1}{|c|}{}                        & Anafi                                                      &\Checkmark         &             &\Checkmark                                                   &\Checkmark  &\Checkmark                                                             &                                                         & 4                                                               & 53                                                      & 25                                                            & 312                                                   & 4                                                        & 4K         & 2018 & 599                                                  \\ \hline
\multicolumn{1}{|c|}{\multirow{3}{*}{Yuneec}} & Mantis Q                                                   &            &\Checkmark          &\Checkmark                                                   &\Checkmark  &\Checkmark                                                             &                                                         & 1.5                                                             & 70                                                      & 33                                                            & 479                                                   &                                                          & 4K         & 2018 & 499                                                  \\ \cline{2-16} 
\multicolumn{1}{|c|}{}                        & \begin{tabular}[c]{@{}c@{}}Typhoon \\ 4K\end{tabular}      &            &\Checkmark          &\Checkmark                                                   &\Checkmark  &\Checkmark                                                             &                                                         & 0.6                                                             & 30                                                      & 25                                                            & 200                                                   & 0.1                                                      & 4K         & 2015 & 499                                                  \\ \cline{2-16} 
\multicolumn{1}{|c|}{}                        & \begin{tabular}[c]{@{}c@{}}Typhoon \\ H Plus\end{tabular}  &            &\Checkmark          &\Checkmark                                                   &\Checkmark  &\Checkmark                                                             &                                                         & 1.2                                                             & 48                                                      & 30                                                            & 1995                                                  & 0.5                                                      & 4K         & 2018 & 1899                                                 \\ \hline
\multicolumn{1}{|c|}{Skydio}                  & R1                                                         &\Checkmark         &             &\Checkmark                                                   &     &                                                                &                                                         & 0.1                                                             & 40                                                      & 16                                                            & 997                                                   & 0.1                                                      & 4K         & 2018 & 1999                                                 \\ \hline
\multicolumn{1}{|c|}{Rabing}                  & Rabing Mini                                                        &\Checkmark         &             &                                                 &     &                                                                &                                                         & 0.1                                                             & 20                                                      & 10                                                            & 90                                                  & 0.1                                                    & 720p         & 2017 & 70                                                 \\ \hline
\multicolumn{1}{|c|}{HASAKEE}                  & H1                                                       &\Checkmark         &             &                                                 &     &                                                                &                                                         & 0.05                                                             & 20                                                      & 7                                                            & 50                                                  & 0.05                                                    & 0.3MP         & 2018 & 83                                                 \\ \hline
\multicolumn{1}{|c|}{Holy Stone}                  & HS190                                                       &         &  \Checkmark           &                                                 &     &                                                                &                                                         & 0.05                                                             & 15                                                      & 7                                                            & 25                                                  & 0.05                                                    & 0.3MP         & 2017 & 35                                                 \\ \hline
\multicolumn{1}{|c|}{TOZO}                  & Q2020                                                       &         & \Checkmark            &                                                 &     &                                                                &                                                         & 0.045                                                             & 12                                                      & 8                                                            & 50                                                  & 0.045                                                    & NO         & 2017 & 60                                                 \\ \hline
\end{tabular}
}
\end{table*}

\subsection{Functionality}
\label{subsection:functionality}

\begin{itemize}
    \item Automatic Maneuvering – Allows the operator to mark a target/trajectory via the GCS, around which the drone will automatically maneuver itself according the given input, based on its GPS device. 
    \item Follow Me – Allows the operator to mark a moving object  via the GCS monitor, which the drone will automatically maneuver itself to follow. This functionality is mainly used by a drone operator that is on the move and cannot control the drone due to the nature of his/her activity (e.g., skiing, bicycling, running, etc.); the follow me functionality is based on computer vision algorithms that detect the moving object by processing the captured video stream.
    \item Return to Home (RTH) – Allows automatic maneuvering of the drone to a predefined home target and landing based on a GPS device. This method is automatically operated when the connection between the drone and its GCS has stopped or been disabled.
    \item Smart Capture – Allows the ability to control the drone using hand gestures captured by the drone's camera during flight. This functionality eliminates the need for using a GCS and is based on computer vision algorithms that interpret hand gestures and translates them to maneuvering commands. This feature is only supported by DJI drones.
\end{itemize}

\subsection{Uses}
\label{subsection:use}
Many individual entrepreneurs, small businesses, large companies, and local authorities have begun to realize the potential of drones and started to adopt drones for various purposes. Regulations are constantly changing in order to support this revolution, allowing drones to fly over populated areas and carry packages \cite{FAA-committee-says-small-drones-should-be-allowed-to-fly-over-cities-and-crowds, President-Trump-Moves-to-Fill-America's-Skies-with-Drones}. Drone use varies according to the drone's capabilities. Table \ref{tab:cmp-drones} presents a comparison of the capabilities of the top selling Mini, Macro, and Nano drones. As can be seen from the table, Mini drones (e.g., DJI Matrice 600 Pro, DJI Inspire 2) can reach a range of up to seven kilometers and support flight times of up to 38 minutes. They are also capable of delivering small packages (e.g., DJI Matrice 600 Pro can carry up to seven kilograms). This type of drone is about to provide an alternative means for delivering a package and many firms launched pilots around the world for delivering goods (e.g., Amazon \cite{Amazon}, UPS FedEx \cite{UPS-FEDEX}), pizza (Dominos \cite{PIZZA-DELIVERY-BY-DRONE-LAUNCHED-BY-DOMINOS}), and emergency healthcare (drugs, blood supply). This type of drones is also used in agriculture (for automatic aerial spraying of liquid pesticides, water, and fertilizers \cite{dji-agriculture}) and by militaries \cite{IDF-DJI,NZ-DJI}. Micro Drones can reach the same flight's range as Mini drones, however they are very limited in their carrying primarily used for professional filming and photography by local authorities (e.g., for disaster management, geographic mapping), law enforcement (e.g., by police, border patrol officers), professional photographers (e.g., media, film industry), and contractors  \cite{drones-applications, top-12-non-military-uses-for-drone, top-38-non-military-uses-for-drone}. The range of Nano drones is limited to 100 meters, and they are considered drones for amateurs. Most of them do not support cutting-edge functionality that is integrated in Mini and Micro drones. Some of them do not even contain an integrated camera. However, a recent development in this area showed that Nano drones can be used for military purposes for targeted assassination in a battlefield \cite{Autonomous-Killer-Drones}.

\begin{table*}[]
\centering
\caption{Threats Mapped to Types of Drones}
\label{tab:threats}
  \resizebox{0.8\linewidth}{!}{%
\begin{tabular}{c|c|c|c|c|c|}
\cline{2-6}
 & \multicolumn{2}{c|}{Privacy} & \begin{tabular}[c]{@{}c@{}}Physical \\Attacks\end{tabular} & Crime & \begin{tabular}[c]{@{}c@{}}Cyber\\ Attacks\end{tabular} \\ \cline{2-6} 
 & \begin{tabular}[c]{@{}c@{}}Video \\ Streaming\end{tabular} & \begin{tabular}[c]{@{}c@{}}Carrying \\Surveillance Equipment\end{tabular} &  &  &  \\ \hline
\multicolumn{1}{|c|}{Nano} & \Checkmark &  & \begin{tabular}[c]{@{}c@{}}Targeted\\assassination \cite{Autonomous-Killer-Drones}\end{tabular} & \begin{tabular}[c]{@{}c@{}}Targeting homes\\for burglaries \cite{Burglarss}\end{tabular}&  \\ \hline
\multicolumn{1}{|c|}{Micro} & \Checkmark &  
\begin{tabular}[c]{@{}c@{}}3D mapping using\\radio transceiver  \cite{karanam20173d}\\ MITM attacks against\\cellular networks  \cite{imsi-catcher} \\ Tracking a person according\\to his/her devices  \cite{Snoopy} \end{tabular}

& \begin{tabular}[c]{@{}c@{}}Carrying radioactive\\sand \cite{Japan-PM}\end{tabular} & \begin{tabular}[c]{@{}c@{}}Smuggling goods\\into prison yards \cite{Big-rise-in-drone-jail-smuggling-incidents}\end{tabular} &  \\ \hline
\multicolumn{1}{|c|}{Mini} & \Checkmark & \Checkmark & \begin{tabular}[c]{@{}c@{}}Carrying a bomb \cite{venezuela}; \\ colliding with an airplane \cite{UAV-Related-Incidents,Gatwick}\end{tabular} & \begin{tabular}[c]{@{}c@{}}Hijacking radio\\
controlled devices \cite{chesaux2014wireless,game-of-drones}\\
Smuggling goods between\\countries \cite{Two-plead-guilty-in-border-drug-smuggling-by-drone,smuggling-iphones,smuggling-tobaco} \end{tabular} & \begin{tabular}[c]{@{}c@{}}Establishing a covert\\channel \cite{nassi2019xerox,guri2017led}\end{tabular} \\ \hline
\multicolumn{1}{|c|}{Swarm} & \Checkmark & \Checkmark & \begin{tabular}[c]{@{}c@{}}Multiple casualty\\ incidents \cite{attack-against-russia}\end{tabular} & Cyber warfare \cite{ronen2017iot} &  \\ \hline
\end{tabular}
}
\end{table*}

\section{Malicious Drone Uses}
\label{Sec:Threat Models}
The current generation of drones provides FPV capabilities that allow operators to fly drones in areas located up to eight kilometers from the operator's location; this can be done both manually and automatically. In addition, modern drones are very small, and they can reach speeds of up to 65 kilometers per hour and carry up to six kilograms. The capabilities that were identified by industry and encouraged this sector to adopt drones for various legitimate purposes \cite{PIZZA-DELIVERY-BY-DRONE-LAUNCHED-BY-DOMINOS,Amazon,CNN-Just-Got-Approved-to-Fly-Drones-Over-Crowds-of-People,top-12-non-military-uses-for-drone} have also been identified by malicious entities that misuse drones for illegitimate purposes. The cutting-edge technology and low price of drones made them accessible to individuals, resulting in an increase in drone sales; this has created new threats and caused the number of drone related incidents to rise significantly in recent years. In this section, we describe the major threats that drones pose to security and privacy today.

\subsection{Spying and Tracking}
The FPV channel provides excellent infrastructure for a malicious operator to spy on people without being detected because: (1) it eliminates the need for a malicious operator to be close to the drone or target by allowing the operator to maneuver the drone from far away to a target that is also far away from the operator's location, (2) it can be secured using encryption, and (3) it supports HD resolutions that enable the attacker to obtain high quality pictures and close-ups (by using the video camera's zooming capabilities) that are captured by the drone, even when the drone is far from the target POI. In addition to the abovementioned, the presence of drones is no longer restricted in populated areas \cite{President-Trump-Moves-to-Fill-America's-Skies-with-Drones,FAA-committee-says-small-drones-should-be-allowed-to-fly-over-cities-and-crowds}. Exploiting these facts, drones have increasingly become a threat to individuals' privacy as evidenced by their use to detect a cheating spouse \cite{cheating}, film random people \cite{26-floor,Woman-grabs-gun-shoots-nosy-neighbour-s-drone} and celebrities \cite{Virginia-Woman-Shoots-Down-Drone-Near-Actor-Robert-Duvalls-Home}, and take intimate pictures of neighbors \cite{not-my-backyard-man-arrested-after-shooting-drone-down}. In addition, empty houses have also been targeted by burglars for robberies using drones \cite{Burglarss}.

Drones can also provide a means of carrying a surveillance device. Several studies have shown that drones equipped with radio transceivers can be used for (1) locating and tracking people across a city \cite{Snoopy} by extracting unencrypted information from the lower layers of the OSI model of radio protocols (e.g., Wi-Fi and Bluetooth) that reveal the MAC addresses of their device's owners (e.g., smartphones, smartwatches) \cite{197130}, (2) 3D through-wall imaging with drones \cite{karanam20173d}, and (3) performing an MITM attack on telephony \cite{imsi-catcher} by downgrading 4G to 2G. Drones can also be used to carry traditional spying devices used to eavesdrop on a conversation (e.g., a laser microphone \cite{laser-microphone}) and perform keylogging (e.g., using a microphone). 

\subsection{Smuggling}
Commercial drones provide optimal infrastructure for smuggling due to their flight range, size, speed, and carrying capabilities. The abovementioned reasons make drone detection very difficult and have caused criminals to adopt drones for smuggling purposes. Drones are currently used for dropping weapons and other contraband into prison yards \cite{Big-rise-in-drone-jail-smuggling-incidents}, and smuggling goods \cite{smuggling-iphones,smuggling-tobaco} and drugs between countries over borders \cite{Two-plead-guilty-in-border-drug-smuggling-by-drone}. In terms of smuggling, using a drone has two major benefits: (1) it eliminates the need for a human smuggler, and (2) even if a smuggling drone is detected and caught, determining the identity of its operator (who might be located a few kilometers from the target) remains a challenge.

\subsection{Physical Attacks}
Exploding and shooting drones are no longer relegated to science fiction. We are now living in an era that was referred to as “terrorism by joystick” by a few sources \cite{terrorism-by-joystick-1,terrorism-by-joystick-2,terrorism-by-joystick-3}. The reasons that led criminals to adopt drones for smuggling, have also led terrorists to adopt drones for various purposes. Just recently, the Venezuelan president was the target of an assassination attempt conducted by two drones while speaking at an event to mark the 81st anniversary of the national army \cite{venezuela}. In 2015, two people were arrested in two different incidents for crashing a DJI Phantom into a tree on the south lawn of the White House \cite{A-Drone-Too-Small-for-Radar-to-Detect-Rattles-the-White-House} and for landing a ‘radioactive' drone on the Japanese Prime Minister's roof \cite{Japan-PM}. These acts have raised many questions about the ability to protect a world leader from an aerial targeted assassination.

Targeting a world leader is not the only threat that drones pose. Drones can cause much greater disasters in terms of the number of casualties by exploding into critical infrastructure. This type of threat was demonstrated by the Greenpeace organization which crashed a Superman shaped drone into a French nuclear plant \cite{greenpeace}. Multiple casualty incidents resulting from explosive drones targeting crowds represent another increasing concern \cite{ISIS, hamas, German-Police}. This threat has been demonstrated in (1) an armed drone attack conducted by unspecified terrorists against Russian military bases in Syria \cite{attack-against-russia}. Commercial airliners are also vulnerable to exploding and colliding drone attacks during takeoff and landing \cite{attacks-against-airplanes}. This threat led to the cancellation of hundreds of flights at Gatwick Airport near London, England, following reports of drone sightings close to the runway \cite{Gatwick}. The recent Gatwick incident is probably the most famous drone airport incident due to the volume of its damage, however there are dozens of reported near miss incidents involving drones all around the world,  and recently such incidents have begun to be reported on weekly and monthly base \cite{UAV-Related-Incidents}.

\subsection{Launching a Cyber Attack}
Cyber-attacks deemed in the past as infeasible due to distance, line of sight, and other factors can now be performed using a drone. Two studies \cite{nassi2019xerox,guri2017led} showed how to establish a covert channel for data infiltration \cite{nassi2019xerox} and exfiltration \cite{guri2017led} to/from an organization. In these studies, the drone was used to carry a transmitter (in \cite{nassi2019xerox}) and a receiver (in \cite{guri2017led}), in order to modulate/demodulate data sent to/from malware installed on an air-gapped network of a target organization. Other studies have shown that a drone equipped with a radio transceiver can be used to hijack a Bluetooth mouse \cite{game-of-drones} to gain access to a wireless office printer \cite{toh2017cyber}; perform wireless spoofing and deauthentication attacks on a targeted user \cite{chesaux2014wireless}; and hijack a Philips Hue smart bulb \cite{ronen2017iot}. Many other known attacks can also be performed from a drone equipped with proper hardware, such as speakers (for triggering smart assistants \cite{zhang2017dolphinattack, carlini2016hidden}) and radio transceivers (for exploiting WPA-2 \cite{vanhoef2017key} and Bluetooth \cite{bihambreaking} vulnerabilities).

\subsection{Summary \& Challenges in the Near Future}
Table \ref{tab:threats} maps threats to types of drones. The potential harm from the abovementioned threats is likely to be amplified in the near future when a malicious operator will be able to operate a swarm of drones simultaneously to perform his/her task, turning a targeted cyber-attack against an organization into cyber warfare and a targeted assassination into a massive terror attack. In addition, many sci-fi scenarios that were unrealistic due to technological limitations are now real. For example, shooting drones have already been documented \cite{shooting-drone}, and they can be used by criminals to rob banks \cite{drone-robber-bank} and individuals \cite{drone-robber-subject}.

\section{Malicious Drone Detection in Restricted Flight Areas }
\label{Sec:detection-restricted}
In this section, we describe methods to detect drones flying over areas that are restricted to flight. Various geofencing methods used to detect the location of a nearby consumer drone have been introduced in the last few years, and the global anti-drone market size is anticipated to reach \$1.85 billion by 2024 \cite{anti-drone-market}. Geofencing methods are very effective at detecting a malicious object in areas that restrict unauthorized entries such as military bases, prisons, etc. These methods can be used to detect drones that are used for purposes of dropping weapons and other contraband into prison yards \cite{Big-rise-in-drone-jail-smuggling-incidents}, smuggling goods and drugs between countries over borders \cite{Two-plead-guilty-in-border-drug-smuggling-by-drone}, and crashing on the White House lawn \cite{A-Drone-Too-Small-for-Radar-to-Detect-Rattles-the-White-House,Secret-Service-Arrests-Man-After-Drone-Flies-Near-White-House}). Some geofencing methods have been adopted from prior military knowledge for detecting UAVs and airplanes. However, as indicated by \cite{eshel2013mobile}, drones are much harder to detect than manned aircraft, so new dedicated methods have been suggested to deal with the challenges that arise from detecting a small, high-speed object flying in the air. The difficulty in detecting drones is an issue modern armies, police departments, and governments are aiming to overcome, as it is a recognized threat to critical infrastructure, operations , and individuals.

\begin{figure}
\centering
\includegraphics[width=0.75\linewidth]{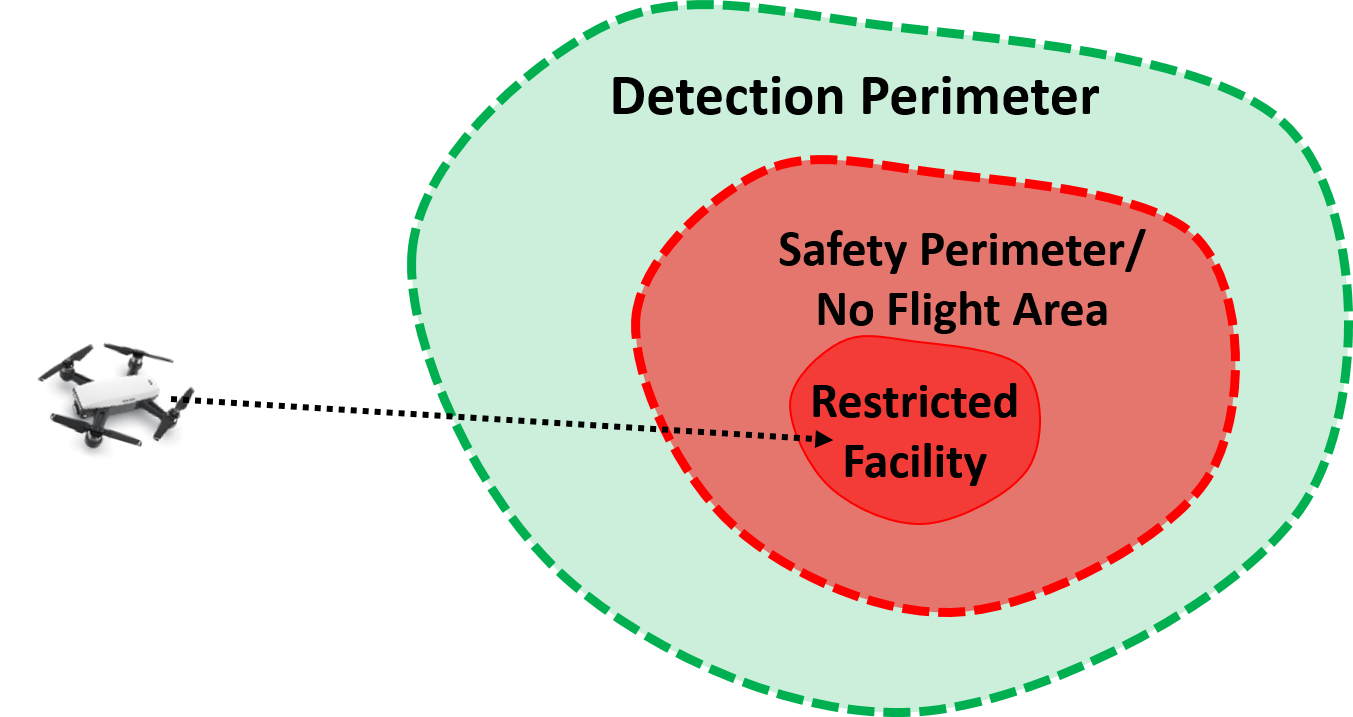}
\caption{Ideal detection scheme for restricted areas.} 
\label{fig:restricted-scheme}
\end{figure}

In this section, we describe dedicated methods used to detect Nano, Micro, and Mini drones. Small and Tactical drone detection methods are not covered here , because the challenges associated with detecting Nano, Micro, and Mini drones are different than those associated with detecting Small and Tactical drones,  and many existing mechanisms for the detection of Small and Tactical drones are ineffective for detecting Nano, Micro, and Mini drones. A number of considerations need to be taken into account when evaluating a detection method. Many factors, including ambient light, weather, false positive rates, ambient noise, cost, line of sight, and detection range influence the effectiveness of each method. In this section, we analyze each method's effectiveness at dealing with issues important for securing a restricted area from the presence of drones. Another question that we address is whether the suggested method can be used for drone identification (i.e., detecting the drone type).

\begin{table*}[htbp]
  \centering
  \caption{Effectiveness of Sensors to Various Environmental Conditions}
\begin{tabular}{c|c|c|c|c|c|c|}
\cline{2-7}
                                                 & \multicolumn{6}{c|}{Sensor's Effectiveness for Detection}                                                                                                                                                                                                                                                                              \\ \cline{2-7} 
                                                 & \multicolumn{2}{c|}{RF}                                                                                                                    & \multicolumn{3}{c|}{Optical}                                                       & Acoustical                                                                \\ \hline
\multicolumn{1}{|c|}{Factor}              & \begin{tabular}[c]{@{}c@{}}Active \\ Radar\end{tabular} & \begin{tabular}[c]{@{}c@{}}RF \\ Scanners\end{tabular}                           & VIS                                                              & IR      & LiDAR &                                                                           \\ \hline
\multicolumn{1}{|c|}{Light}                      &\Checkmark                                                      &\Checkmark                                                                               &\Checkmark                                                               &         &\Checkmark    &\Checkmark                                                                        \\ \hline
\multicolumn{1}{|c|}{Darkness}                   &\Checkmark                                                      &\Checkmark                                                                               &                                                                  &\Checkmark      &\Checkmark    &\Checkmark                                                                        \\ \hline
\multicolumn{1}{|c|}{Noise}                      &\Checkmark                                                      &\Checkmark                                                                               &\Checkmark                                                               &\Checkmark      &\Checkmark    &                                                                           \\ \hline
\multicolumn{1}{|c|}{Birds}                      &                                                         &\Checkmark                                                                               &                                                                  &         &       &\Checkmark                                                                        \\ \hline
\multicolumn{1}{|c|}{Adverse Weather Conditions} &                                                         &                                                                                  &                                                                  &         &       &                                                                           \\ \hline
\multicolumn{1}{|c|}{Drone Identification}       &                                                         &\Checkmark                                                                               &\Checkmark                                                               & Limited &       &\Checkmark                                                                        \\ \hline
\multicolumn{1}{|c|}{Autonomous Drone Detection}  &                                                         &                                                                                  &\Checkmark                                                               &\Checkmark      &\Checkmark    &\Checkmark                                                                        \\ \hline
\multicolumn{1}{|c|}{Multiple Drone Detection}   &\Checkmark                                                      & \begin{tabular}[c]{@{}c@{}}Only if drones use \\ different channels\end{tabular} &\Checkmark                                                               &\Checkmark      &\Checkmark    & \begin{tabular}[c]{@{}c@{}}Only different \\ types of drones\end{tabular} \\ \hline
\multicolumn{1}{|c|}{Cost}                      &                                                         &\Checkmark                                                                               &\Checkmark                                                               &         &       &\Checkmark                                                                        \\ \hline
\multicolumn{1}{|c|}{Long Range Detection}       &\Checkmark                                                      &\Checkmark                                                                               & \begin{tabular}[c]{@{}c@{}}Require \\ focusing lens\end{tabular} &         &\Checkmark    &                                                                           \\ \hline
\multicolumn{1}{|c|}{Immunity to NLOS}           &                                                         &\Checkmark                                                                               &                                                                  &         &       &\Checkmark                                                                        \\ \hline
\multicolumn{1}{|c|}{Locating}                   &\Checkmark                                                      & Multiple                                                                         &\Checkmark                                                               &\Checkmark      &\Checkmark    & Multiple                                                                  \\ \hline
\end{tabular}
  \label{tab:methods}%
\end{table*}%

We consider facilities such as airports, jails, and military bases as restricted facilities. They are considered sensitive or dangerous due to the nature of the activity performed at the facility and the danger posed from physical and cyber-attacks against them. These kinds of facilities are largely isolated from urban environments and considered no-fly areas for drones. Ideally, drone flights are prohibited from a wider perimeter beyond the restricted area; we consider this area a safety perimeter, however it may vary according to physical limitations, local aviation authorities, and regulations. A safety perimeter is required to prevent attackers from spying and triggering a cyber-attack against the restricted area. Beyond the safety perimeter, there is also a detection perimeter in which drones are allowed to fly; the area within the detection perimeter must be monitored so that a flying  drone that enters the safety perimeter is detected early enough. The ideal detection scheme is presented in Figure \ref{fig:restricted-scheme}. 

Commercial drones are programmed to avoid entering a no-fly area  by automatically landing on the ground or returning to a user-defined home coordinate before entering such a zone. They contain an internal database of no-fly areas defined by polygons of GPS coordinates. The functionality that prevents drones from flying over no-fly areas is extremely important and prevents operators from accidentally entering a no-fly area, however malicious operators can bypass this functionality using counter mechanisms that are sold online.

\subsection{Radar}
Monostatic radar (transmitter and receiver are collocated) is a traditional method of drone detection which detects the electromagnetic waves (EM) reflected from objects in order to determine a drone's range, speed, and velocity. 
However, the detection of the smallest consumer drones requires high-frequency radar systems, since drones reflect in a frequency of 10Ghz \cite{10GHz-radar}, and can evade low-frequency systems \cite{A-Drone-Too-Small-for-Radar-to-Detect-Rattles-the-White-House}. Several studies analyzed monostatic radar (35 GHz \cite{drozdowicz201635} and 9.4 GHz \cite{de2016flight}) to detect the distance of a nearby drone. Another study \cite{molchanov2014classification} showed that distinguishing between a drone and a bird can done using machine learning algorithms, by extracting features from micro-Doppler signatures obtained at a frequency of 9.5 GHz. Several methods suggested the use of bistatic radar (transmitter and receiver are not collocated) by analyzing RF signals in order to detect the presence of a drone. A recent study \cite{liu2017digital} showed how the trail of a DJI Phantom 4 can be detected from an antenna by analyzing TV signals transmitted from an existing TV tower, while another method suggested using multiple antennas to receive and process the orthogonal frequency-division multiplexing (OFDM) echoes of UAV, which are originally transmitted by the nearby base stations. Another study suggested the use of a radar \cite{hoffmann2016micro} by triangulating data obtained from three receivers located linearly on a plane, 50 meters from each other (where the middle radar is the transceiver), in order to locate drone in space.

Radar can be used to detect and locate a drone in space. However, radar detection can be unreliable, as adverse weather conditions affect the reflected wavelength, distorting the wave. Radar also fails to identify the drone type and suffers from high false positive rates, since it cannot distinguish between birds and drones. In addition, the type of radar that is needed to detect drones is expensive. Due to the abovementioned limitations only 13 out of 33 companies use  radar\cite{DroneDetectionSystem,GROK,Gamekeeper16U,UAVX,LightWave,AUDS,HARRIER,Skylight,SharpEve,ELVIRA,Giraffe,A2000,SQUIRE}, and eight of them \cite{DroneDetectionSystem,UAVX,AUDS,HARRIER,Skylight,SharpEve,ELVIRA,SQUIRE} use radar in the first stage of detection and another type of sensor in the second stage.

\subsection{RF Scanner \& Spectrum Analyzer}
RF scanners and spectrum analyzers are primarily used to detect drone radio signatures by detecting bands that are known to be used by drones and other radio signatures. They can be used for (1) classifying a suspicious transmission as an FPV channel, and (2) locating a drone in space. One study analyzed \cite{birnbach2017wi}  the received signal strength indication (RSSI) patterns of Wi-Fi signals for the detection of approaching, escaping, and spying Wi-Fi drones. This method can be applied using a simple Wi-Fi receiver, however it is only effective when a line of sight between the Wi-Fi receiver and drone exists, and its accuracy for detecting a drone among other moving IoT devices that transmit Wi-Fi signals (e.g., smartwatch, smartphone) was not validated. Another study \cite{nguyen2017matthan, nguyen2018cost} analyzed 10 seconds of RF signals captured by SDR and found that the RF signatures of commercial Wi-Fi drones can be (1) detected with high accuracy (84.9\%) from a distance of 600 meters, and (2) used to identify the detected drone type with variable accuracy (64-89\%) depending on the drone. Another study \cite{nguyen2016investigating} used USRP to detect drones and found that the presence of the drone can also  be detected in an urban environment, however it is challenging to observe a drone's RF signature when the drone more than 50 meters away without increasing the gain  of the receiver's antenna panel. Two studies \cite{scheller2017detecting,shi2017detection} used machine learning algorithms to classify drone transmissions. Another study \cite{peacock2013towards} suggested detecting the presence of a drone by analyzing the MAC addresses of known drones of nearby captured access points . However, attackers can evade the suggested detection method by changing a drone's MAC address. RF scanners can be very effective at detecting the presence of a drone and identifying its type by comparing them  to known used bands, however RF scanners suffer from an inability to accurately locate a drone in space unless they are triangulated  \cite{mototolea2018detection}. In addition, attackers can evade detection by using drones that transmit on a dedicated band that is not popular for FPV use.

\begin{table*}[]
\caption{Characteristics of Commercial Devices for Drone Detection }
  \resizebox{1.0\linewidth}{!}{%
\begin{tabular}{cc|c|c|c|c|c|c|c|c|c|c|c|c|}
\cline{3-14}
                                                                                             &                                                                                                                & \multicolumn{2}{c|}{Radio}                                    & \multicolumn{4}{c|}{Optical}                                                                  & Acoustic & \multicolumn{5}{c|}{Features}                                                                                    \\ \hline
\multicolumn{1}{|c|}{Company Name}                                                           & Product Name                                                                                                   & Radar & \begin{tabular}[c]{@{}c@{}}RF \\ Scanner\end{tabular} & Camera & LiDAR & \begin{tabular}[c]{@{}c@{}}Electro-\\ optical\\ Camera\end{tabular} & Thermal & Microphone & \begin{tabular}[c]{@{}c@{}}Detection\\ Range \\ (KM)\end{tabular} & Identification & Angle & Locating & Mobility \\ \hline
\multicolumn{1}{|c|}{3DEO}                                                                   & \begin{tabular}[c]{@{}c@{}}Rogue Drone \\ Detection \& \\  Mitigation \cite{RogueDrone}\end{tabular}           &       &                                                       &        &\Checkmark    &                                                                    &         &            & 2                                                                 &                &       &\Checkmark       &          \\ \hline
\multicolumn{1}{|c|}{Aaronia}                                                                & \begin{tabular}[c]{@{}c@{}}Drone Detection \\ System \cite{DroneDetectionSystem}\end{tabular}                  &\Checkmark    &                                                       &\Checkmark     &       &                                                                    &\Checkmark      &            & 50                                                                &\Checkmark             &       &\Checkmark       &\Checkmark       \\ \hline
\multicolumn{1}{|c|}{\multirow{2}{*}{Anti-Drone.eu}}                                         & GROK \cite{GROK}                                                                                               &\Checkmark    &                                                       &        &       &                                                                    &         &            & 4                                                                 &\Checkmark             &       &\Checkmark       &          \\ \cline{2-14} 
\multicolumn{1}{|c|}{}                                                                       & Droneshield \cite{droneShield}                                                                                 &       &                                                       &        &       &                                                                    &         &\Checkmark         & 0.5                                                               &                &       &          &          \\ \hline
\multicolumn{1}{|c|}{Aveillant}                                                              & \begin{tabular}[c]{@{}c@{}}Gamekeeper 16U - \\ Holographic Radar \cite{Gamekeeper16U}\end{tabular}             &\Checkmark    &                                                       &        &       &                                                                    &         &            & 5                                                                 &                &       &\Checkmark       &          \\ \hline
\multicolumn{1}{|c|}{Black Sage - BST}                                                       & UAVX \cite{UAVX}                                                                                               &\Checkmark    &                                                       &\Checkmark     &       &                                                                    &\Checkmark      &            & 0.5                                                               &                & 90    &\Checkmark       &\Checkmark       \\ \hline
\multicolumn{1}{|c|}{C speed LLC}                                                            & LightWave Radar \cite{LightWave}                                                                               &\Checkmark    &                                                       &        &       &                                                                    &         &            &                                                                   &                &       &\Checkmark       &          \\ \hline
\multicolumn{1}{|c|}{CACI}                                                                   & SkyTracker \cite{SkyTracker}                                                                                   &       &\Checkmark                                                    &        &       &                                                                    &         &            &                                                                   &\Checkmark             &       &          &          \\ \hline
\multicolumn{1}{|c|}{CerbAir}                                                                & DroneWatch \cite{DroneWatch}                                                                                   &       &\Checkmark                                                    &        &       &                                                                    &         &            & 1                                                                 &                &       &\Checkmark       &\Checkmark       \\ \hline
\multicolumn{1}{|c|}{\begin{tabular}[c]{@{}c@{}}Chess Dynamics\\ Ltd\end{tabular}}                                                     & AUDS \cite{AUDS}                                                                                               &\Checkmark    &                                                       &        &       &\Checkmark                                                                 &\Checkmark      &            & 10                                                                &                & 180   &\Checkmark       &\Checkmark       \\ \hline
\multicolumn{1}{|c|}{DeDrone.com}                                                            & DroneTracker \cite{DroneTracker}                                                                               &       &\Checkmark                                                    &\Checkmark     &       &                                                                    &         &            &                                                                   &\Checkmark             &       &          &\Checkmark       \\ \hline
\multicolumn{1}{|c|}{\multirow{2}{*}{DeTect}}                                                & DroneWatcher \cite{DroneWatcher}                                                                               &       &\Checkmark                                                    &        &       &                                                                    &         &            & 1.6-3.2                                                           &\Checkmark             &       &          &\Checkmark       \\ \cline{2-14} 
\multicolumn{1}{|c|}{}                                                                       & HARRIER DSR \cite{HARRIER}                                                                                     &\Checkmark    &                                                       &\Checkmark     &       &                                                                    &         &\Checkmark         & 3.2                                                               &\Checkmark             &       &\Checkmark       &          \\ \hline
\multicolumn{1}{|c|}{\begin{tabular}[c]{@{}c@{}}Digital \\ Global \\ Systems\end{tabular}}   & SigBASE \cite{SigBASE}                                                                                         &       &\Checkmark                                                    &        &       &                                                                    &         &            &                                                                   &                &       &          &\Checkmark       \\ \hline
\multicolumn{1}{|c|}{DroneShield}                                                            & \begin{tabular}[c]{@{}c@{}}FarAlert/WideAlet \\ Sensors \cite{FarAlert}\end{tabular}                           &       &                                                       &        &       &                                                                    &\Checkmark      &\Checkmark         & 1                                                                 &                & 30    &          &\Checkmark       \\ \hline
\multicolumn{1}{|c|}{Gryphon Sensors}                                                        & Skylight \cite{Skylight}                                                                                       &\Checkmark    &\Checkmark                                                    &        &       &\Checkmark                                                                 &\Checkmark      &            & 3-10                                                              &                & 360   &\Checkmark       &\Checkmark       \\ \hline
\multicolumn{1}{|c|}{\begin{tabular}[c]{@{}c@{}}HGH \\ Infrared \\ Systems\end{tabular}}     & \begin{tabular}[c]{@{}c@{}}UAV Detection \\ \& Tracking \cite{UAVdetection}\end{tabular}                       &       &                                                       &\Checkmark     &\Checkmark    &                                                                    &\Checkmark      &            &                                                                   &                & 360   &          &          \\ \hline
\multicolumn{1}{|c|}{\begin{tabular}[c]{@{}c@{}}Kelvin Hughes\\ Limited\end{tabular}}                                                  & {\begin{tabular}[c]{@{}c@{}}SharpEve SxV \\Radar \cite{SharpEve}    \end{tabular}}                                                                         &\Checkmark    &                                                       &\Checkmark     &       &                                                                    &\Checkmark      &            & 1.5                                                               &                & 360   &\Checkmark       &\Checkmark       \\ \hline
\multicolumn{1}{|c|}{MAGNA}                                                                  & Drone Detection \cite{DroneDetection2}                                                                         &       &                                                       &\Checkmark     &       &                                                                    &\Checkmark      &\Checkmark         & 0.5-1                                                             &                &       &          &          \\ \hline
\multicolumn{1}{|c|}{Microflown AVISA}                                                       & Skysentry AMMS \cite{Skysentry}                                                                                &       &\Checkmark                                                    &        &       &                                                                    &         &\Checkmark         & 0.4-1                                                             &                & 360   &\Checkmark       &          \\ \hline
\multicolumn{1}{|c|}{Mistral Solutions}                                                      & \begin{tabular}[c]{@{}c@{}}Drone Detection \\ and Classification \\ System \cite{DroneDetection1}\end{tabular} &       &\Checkmark                                                    &\Checkmark     &       &                                                                    &\Checkmark      &            & 1                                                                 &\Checkmark             &       &          &          \\ \hline
\multicolumn{1}{|c|}{ORELIA}                                                                 & Drone-Detector \cite{DroneDetector}                                                                            &       &                                                       &        &       &                                                                    &         &\Checkmark         & 0.1                                                               &                & 360   &          &          \\ \hline
\multicolumn{1}{|c|}{Quanergy Systems}                                                       & \begin{tabular}[c]{@{}c@{}}Q-Guard - \\ LiDar X-Drone \cite{Qguard}\end{tabular}                               &       &                                                       &        &\Checkmark    &                                                                    &         &            & 0.1                                                               &                &       &          &          \\ \hline
\multicolumn{1}{|c|}{Rinicom}                                                                & SKY PATRIOT \cite{SkyPATRIOT}                                                                                  &       &                                                       &\Checkmark     &       &\Checkmark                                                                 &\Checkmark      &            & 0.8                                                               &\Checkmark             &       &          &          \\ \hline
\multicolumn{1}{|c|}{\begin{tabular}[c]{@{}c@{}}Rinicom and \\ METIS Aerospace\end{tabular}} & SKYPERION \cite{SkyPERION}                                                                                     &       &\Checkmark                                                    &        &       &                                                                    &         &\Checkmark         &                                                                   &                &       &          &          \\ \hline
\multicolumn{1}{|c|}{\begin{tabular}[c]{@{}c@{}}ROBIN \\ Radar \\ Systems\end{tabular}}      & ELVIRA \cite{ELVIRA}                                                                                           &\Checkmark    &                                                       &\Checkmark     &       &                                                                    &         &            &                                                                   &                &       &\Checkmark       &\Checkmark       \\ \hline
\multicolumn{1}{|c|}{\begin{tabular}[c]{@{}c@{}}Rohde and \\ Schwarz\end{tabular}}           & {\begin{tabular}[c]{@{}c@{}} R\&S \\ARDRONIS-I \cite{ARDRONIS}   \end{tabular}}                                                                           &       &\Checkmark                                                    &        &       &                                                                    &         &            & 1-2                                                               &\Checkmark             &       &          &          \\ \hline
\multicolumn{1}{|c|}{SAAB Group}                                                             & \begin{tabular}[c]{@{}c@{}}Giraffe AMB \\ Radar - ELSS \cite{Giraffe}\end{tabular}                             &\Checkmark    &                                                       &        &       &                                                                    &         &            & 30-470                                                            &                & 360   &\Checkmark       &\Checkmark       \\ \hline
\multicolumn{1}{|c|}{Sensofusion}                                                            & AIRFENCE \cite{AIRFENCE}                                                                                       &       &\Checkmark                                                    &        &       &                                                                    &         &            &                                                                   &                &       &          &\Checkmark       \\ \hline
\multicolumn{1}{|c|}{SpotterRF}                                                              & \begin{tabular}[c]{@{}c@{}}A2000 Radar \\ UAVX \cite{A2000}\end{tabular}                                       &\Checkmark    &                                                       &        &       &                                                                    &         &            & 0.2-1                                                             &                & 45/90 &\Checkmark       &\Checkmark       \\ \hline
\multicolumn{1}{|c|}{\begin{tabular}[c]{@{}c@{}}Squarehead \\ Technology\end{tabular}}       & DiscovAir \cite{DiscovAir}                                                                                     &       &                                                       &        &       &                                                                    &         &\Checkmark         &                                                                   &                &       &          &          \\ \hline
\multicolumn{1}{|c|}{\begin{tabular}[c]{@{}c@{}}TCI \\ International\end{tabular}}           & BlackBird \cite{BlackBird}                                                                                     &       &\Checkmark                                                    &        &       &                                                                    &         &            &                                                                   &                &       &          &\Checkmark       \\ \hline
\multicolumn{1}{|c|}{Thales}                                                                 & SQUIRE \cite{SQUIRE}                                                                                           &\Checkmark    &                                                       &        &       &                                                                    &\Checkmark      &            & 48                                                                &                &       &\Checkmark       &\Checkmark       \\ \hline
\end{tabular}
\label{tab:industry}%
}
\end{table*}

\subsection{Optical}
Based on cameras that detect visible frequencies , several studies suggested methods to detect a drone and its trajectory from a single video stream by detecting motion cues \cite{hu2017detection,rozantsev2015flying}, visual marks \cite{santana2014trajectory}, and shape descriptors \cite{unlu2018using}. Other studies trained a neural network \cite{saqib2017study,aker2017using,unlu2018generic} or used multiple fixed ground cameras \cite{rozantsev2017flight} for the same purpose. While the abovementioned methods can also be used to accurately locate and identify (using a preliminary database) drones, they suffer from false positive detections due to the similarities between the movements of drones and birds. They also suffer from high false negative rates due to the increasing number of drone models, the use of non-commercial drones, and ambient darkness.

In order to address the compromised drone detection rate in dark conditions, several studies suggested using thermal cameras that capture invisible wavelengths. A recent study \cite{muller2017robust} suggested using short-wave infrared (SWIR) for night detection. Another study \cite{birch2017counter} performed a comparison of drone detection at various distances using short-wave infrared (SWIR), mid-wave infrared (MWIR), and long-wave infrared (LWIR) imagers and found that (1) SWIR imagers do not appear to be good candidates for drone detection due to their reliance on external light sources, sensitivity to the SWIR absorbing materials used on UAS, and frequent capturing of fast moving and bright insects; (2) LWIR imaging may be best suited
for the detection and assessment of drones, however nuisance sources such as birds will also be captured in the LWIR, and certain angles above the horizon may present challenges due to the temperature equivalence of the target and the background; and (3) MWIR may offer better clutter rejection, while still relying upon self-emitted photons from targets. A recent study \cite{church2018aerial} analyzed the detection of drones using a LiDAR sensor and found that (1) commercial drone speed does not affect the accuracy of detection, and (2) drones can be detected from a distance of a few hundred meters. The greatest disadvantage associated with using an infrared camera and LiDAR is their inability to identify a drone due to the low resolution of captured images.

Typically, cameras that capture visible and invisible wavelengths are combined to support detection throughout the day and night. In addition, 16 companies  \cite{RogueDrone,DroneDetectionSystem,UAVX,AUDS,DroneTracker,HARRIER,FarAlert,Skylight,UAVdetection,SharpEve,DroneDetection2,DroneDetection1,Qguard,SkyPATRIOT,ELVIRA,SQUIRE} out of 33 companies use at least a single optical sensor to detect drones, including electro-optical sensors which have not been suggested in academic studies for the use of identifying the range of a detected drone. However, both capturing visible and invisible optical sensors  (1) are dependent on the line of sight, (2) require excellent focus capabilities to detect drones located far from the camera, (3) require multiple cameras or a 360 camera to provide full 360 degree detection, and (4) are not immune to adverse weather conditions.

\subsection{Acoustic}
Acoustic detection methods are not dependent on the line of sight or the size of the target UAV , and many studies have suggested the use of a microphone array to detect drones by analyzing the noise of the rotors. Other studies presented methods used to detect drones based on comparing a drone's captured acoustic signature with other signatures stored in a database of previously collected sound signatures. A recent study \cite{kim2017real} compared a drone's FFT signal obtained from a microphone using machine learning techniques, while other research \cite{mezei2016drone} used correlation  for comparison. Acoustic signature collection is a major issue for acoustic detection , however factors such as wind, temperature, time of day, obstacles, and other sounds can bend the sound waves, changing the direction the sound will travel (\cite{mirelli2009statistical}). The collection of a sound signal on a hot day with little wind on an open plain will be significantly different than collection of the signal on a cold, windy night in a forest \cite{mirelli2009statistical,roseveare2006robust}. Several studies suggested methods that triangulate sound obtained from centralized \cite{chang2018surveillance} and distributed \cite{sedunov2016passive} microphone arrays in order to detect a drone's direction of arrival and location. Another study \cite{yakubovskiy2015feature} suggested a two-layer feature extractor that can be used to detect drones. While acoustic methods can be used to identify a drone and locate its presence (using multiple distributed microphones ), relying on acoustic signature methods for drone detection suffers from false negative detections due to the increasing number of drone models and false positive detections due to ambient noise. In addition, these methods are very limited in their ability to detect a drone from a distance and require the application of noise filtering techniques and calibration for different environments \cite{case2008low}. The abovementioned limitations are likely the reason that microphones are only used by eight out of 33  companies for drone detection.

\subsection{Hybrid}
 Table \ref{tab:methods} presents an analysis of sensors' ability to meet drone detection system requirements. As can be seen, there is no sensor capable of fulfilling all of the requirements, confirming the findings of \cite{hommes2016detection,laurenzis2017multi}. In order to overcome the limitations that arise from using a single sensor, many sensor fusion methods have been suggested. Several methods \cite{liu2017drone, busset2015detection, christnacher2016optical} pproposed combining acoustic and optical methods, using (1) an acoustic camera, or (2) a camera with a microphone array. While in \cite{liu2017drone, busset2015detection} an optical sensor was used along with an acoustic sensor, in \cite{christnacher2016optical} a moving optical sensor was positioned in the center of the secured area, and several microphones were located outside of the area to provide two stages of detection: direction of arrival (using a microphone) and location (using a camera). Other studies suggested combining optical and RF methods such as LiDAR and radar \cite{de2016flight}. Other research proposed the use of all three methods (acoustic, optical, and RF) in order to detect and locate a drone in space \cite{shi2011detecting, hommes2016detection, vasquez2008multisensor}. In general, using more than a single sensor is the approach taken by 16 of 33  companies for drone detection, however the main disadvantage of the multiple sensor approach is the total cost of applying the method.
 
\subsection{Other Methods}
Other novel methods have been suggested for drone detection. In \cite{boddhu2013collaborative}, the authors suggested using "humans and sensors" in order to create a collaborative network for detecting drones. Their approach relies on people who use their smartphones to capture photos of detected drones and send this information to a centralized server. However, this approach is not practical for real use cases.

\subsection{Drone Detection Industry}
Table \ref{tab:industry} provides a comparison of the 33 largest drone detection companies \cite{game-of-drones} that sell commercial devices for drone detection, comparing them based on the sensors they use and the features they support according to their specifications. As can be seen, 24 companies stick with traditional radio methods (using radar or a RF scanner), 16 companies use some type of optical sensor, and only eight companies use acoustic methods. Eighteen companies use multiple sensors as a means of detection . Only nine companies use sensors to identify detected drones. We made the following interesting observations regarding drone detection mechanisms:

\begin{itemize}
\item \textbf{Asymmetric costs} - Some of these technologies are sold for millions of dollars \cite{irondome-5-million} and are used to detect drones that can be purchased for hundreds of dollars. In addition, the financial damage that can be caused by a drone that costs a few hundreds of dollars is  greater (by several orders of magnitude) than the price of a drone, as was demonstrated in the Gatwick Airport drone incident \cite{Gatwick} that affected about 140,000 passengers and over 1,000 flights and other incidents.
\item \textbf{Different use cases require different solutions} - Factors like detection range and mobility are most likely to determine the chosen solution.
\item \textbf{Variety of sizes} - Some detection mechanisms are the size of a suitcase, while others are much larger (half the size of a car) and require a number of operators.
\item \textbf{Modular products} - Their systems can work based on a single sensor but also support adding additional sensors. 
\end{itemize}

However, as was proven in many cases, detecting a drone is a major challenge, even for commercial drones whose radio, acoustic, and visual signatures  are published or can be learned. An attacker that would like to exploit this fact  to evade drone detection mechanisms may try to use a drone with a different signature by building his/her own drone or changing a drone's signatures (visual and radio).


\section{Malicious Drone Detection in Non-Restricted Flight Areas}
\label{sec:detection-non-restricted}

\begin{figure*}
\centering
\includegraphics[width=0.36\textwidth]{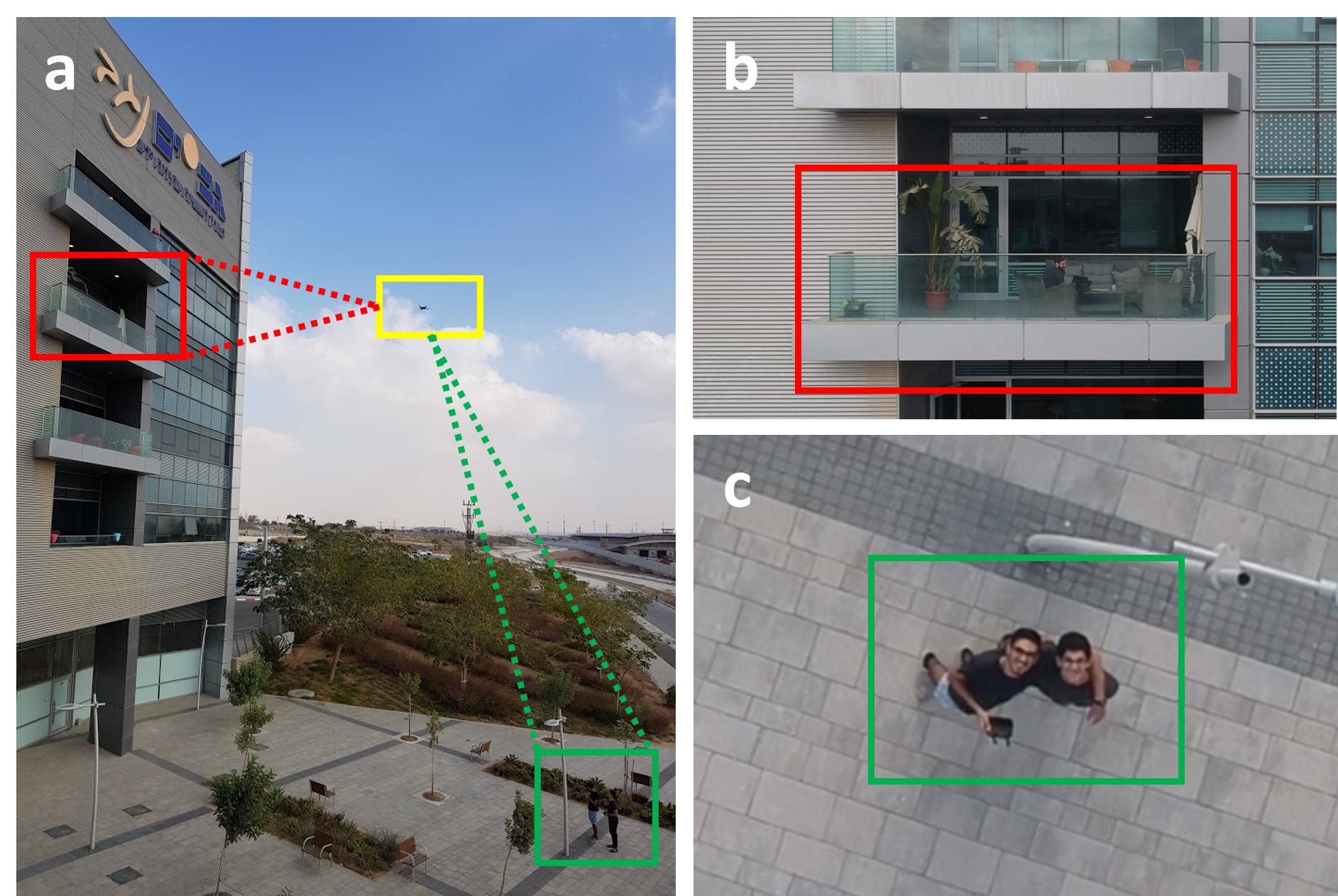}
\includegraphics[width=0.43\textwidth]{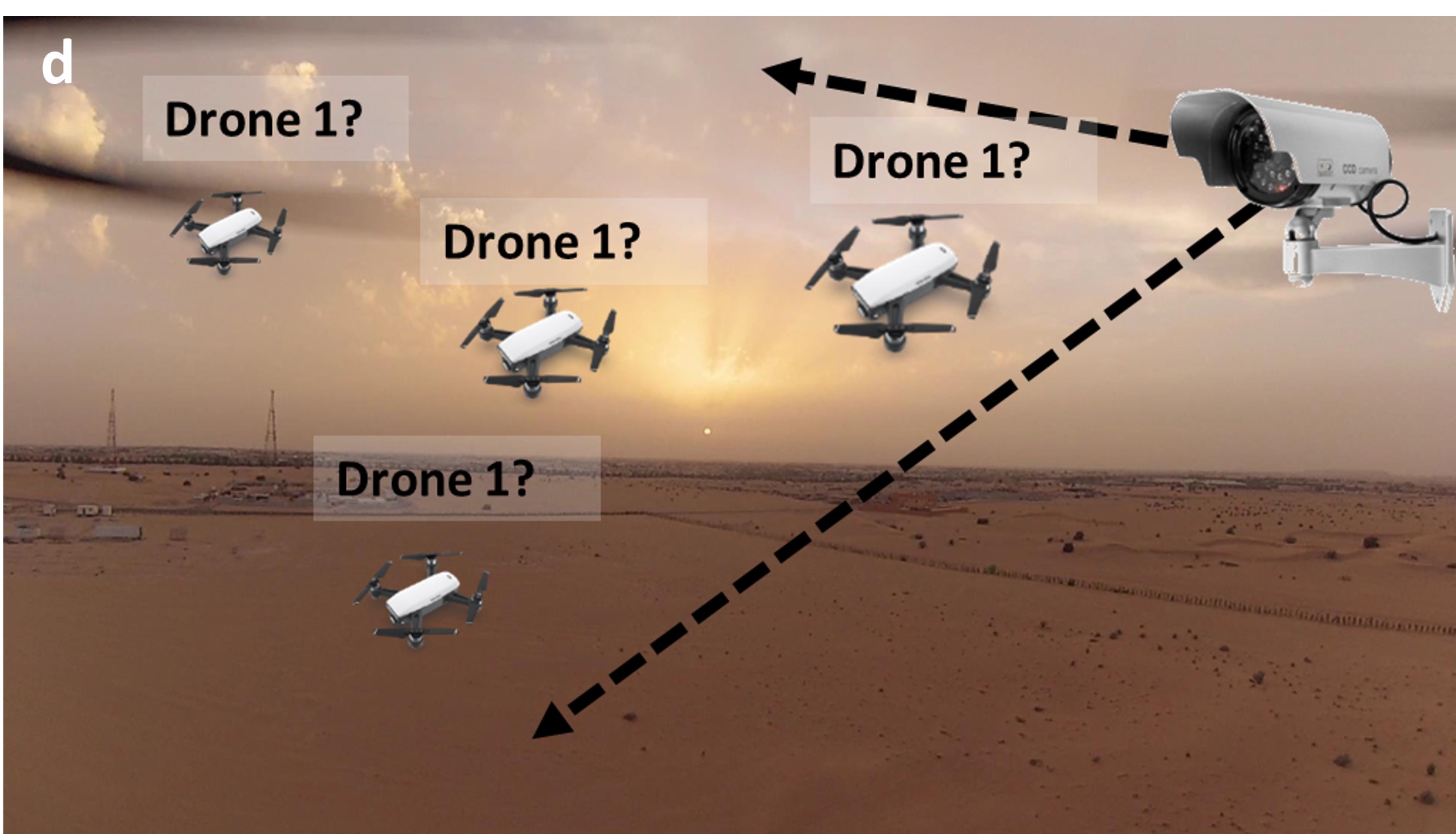}
\caption{Spying classification problem - legitimate and illegitimate use of a drone from the same location: (a) a drone boxed in yellow, two people boxed in green, and the window of an organization boxed in red, (b) illegitimate use of the drone camera to film the organization, (c) legitimate use of the drone for selfie purposes. Identification problem - when all drones look identical (d), it is difficult to match virtual IDs to physical drones in the camera view.} 
\label{fig:problems}
\end{figure*}

In no-flight  zones, a drone is considered dangerous/malicious based its location, so systems that are aimed at detecting and locating a drone (e.g., radar, LiDAR, etc.) can provide a solution. However, detecting a malicious drone in non-restricted areas remains a challenge \cite{nassi2018game,nassi-smashing}. Many countries around the world allow drones to fly in populated/urban areas \cite{President-Trump-Moves-to-Fill-America's-Skies-with-Drones, FAA-committee-says-small-drones-should-be-allowed-to-fly-over-cities-and-crowds} as part of a new "open skies" policy. This policy has encouraged an increasing number of organizations to adopt drones for various legitimate purposes. Drones are currently used for pizza delivery \cite{PIZZA-DELIVERY-BY-DRONE-LAUNCHED-BY-DOMINOS}, the shipment of goods \cite{Amazon}, filming \cite{CNN-Just-Got-Approved-to-Fly-Drones-Over-Crowds-of-People}, and many other legitimate purposes \cite{top-12-non-military-uses-for-drone}.  Allowing drones to fly in populated areas has created two main challenges : purpose detection problem  and identification problem. Given that we are living in an era in which drones are flying among us, we argue that there is major scientific gap in the area of malicious drone detection in non-restricted flight areas, particularly in light of the challenges  that have arisen as a result of the increased use and freedom of drones.

\subsection{Purpose Detection Problem}
The greatest challenge stemming from allowing drones to fly in populated areas is detecting the drone's purpose. This problem has significant impact in several areas, the best-known of which is privacy: drones are being used by their operators to conduct privacy invasion attacks, in which they spy on people and other targets \cite{cheating,26-floor,Woman-grabs-gun-shoots-nosy-neighbour-s-drone,Virginia-Woman-Shoots-Down-Drone-Near-Actor-Robert-Duvalls-Home,not-my-backyard-man-arrested-after-shooting-drone-down}.

More specifically, given a drone that is passing near a house,\textbf{how can we tell whether the drone is being used for a legitimate purpose (e.g., delivering pizza) or an illegitimate purpose (e.g., taking a peek at a person showering in his/her own house)?} The use of traditional geofencing methods (e.g., radar, LiDAR) as a means of detecting a privacy invasion attack in non-restricted areas (e.g., residential neighborhoods) will fail to distinguish between the legitimate use of a nearby drone and illegitimate use that invades an individual's privacy. This distinction can only be made based on the orientation of the drone's video camera rather than on the drone's location; more specifically, differentiation between illegitimate and legitimate use of a drone can only be accomplished by determining the exact POI (point of interest) being streamed over the video channel and not according to the drone's location, as demonstrated in Figure \ref{fig:problems}.

A recent study from NDSS 2017 \cite{birnbach2017wi} suggested a method for detecting a privacy invasion attack based on a drone's flight behavior. They observed that a privacy invasion attack conducted by an operator that is located far from the target (e.g., one kilometer) consists of three unique flying episodes : approaching the target, spying on the target, and escaping from the target. Based on this observation, they suggested detecting the correlative radio patterns from a static radio receiver by analyzing the RSSI of its  FPV channel. However, as was indicated in a recent study from S\&P 2019 \cite{dronescrypta}, this approach cannot be used to detect whether a specific POI is being filmed, as demonstrated in Figure \ref{fig:problems}a.

In order to detect whether a specific POI is being filmed, a recent study from S\&P 2019 suggested a cryptanalysis approach \cite{nassi2018game,nassi-smashing}. The authors found that they could influence the number of transmitted packets being sent from a Wi-Fi FPV drone (e.g., DJI Spark, Parrot Bebop 2, Parrot Anafi, etc.) over its FPV channel by using an LED flicker and exploiting the drone's video compression process. They showed that with each on/off change of a flicker at a given frequency, a larger amount of data is sent over the FPV channel, causing a watermark to be added to the FPV channel that can be detected by analyzing the frequency \cite{dronescrypta} and time \cite{nassi2018game} domains via a network interface card possessed by static \footnote{\label{window} \url{https://youtu.be/4icQwducz68}} and moving targets \footnote{\label{car} \url{https://youtu.be/9PVaDpMsyQE}}.

However, currently the suggested cryptanalysis approach is limited to Wi-Fi FPV drones and can be bypassed by eliminating the video compression stage or by using two video cameras. Currently, no other methods have been suggested to determine whether a passing drone is being used for a legitimate/illegitimate purpose, and as mentioned earlier, there is a major scientific gap in this area.


\subsection{Identification Problem}
The second problem created by allowing drones to fly in populated areas is referred to as the identification problem \cite{Ruiz:2018:IRD:3236498.3214283}. Identifying a drone, i.e., determining the association between drones' physical IDs (such as the drone's MAC address) and visual IDs (such as object tracker output) is an unresolved issue with major impact. For example, given several drones hovering near each other that were captured by a video camera and radio receiver (e.g., radar, spectrum analyzer), h\textbf{how can we distinguish between a foe drone (used to carry a bomb or to spy) and a friend drone (used by a farmer to fertilize its corps) in cases which both drones are identical?} Even if the detected drones use a radio-based authentication protocol (e.g., transmitting their MAC IDs) to identify themselves, we cannot determine a foe from a friend. This problem is illustrated in Figure \ref{fig:problems}d.

Existing "friend or foe" \cite{friend-or-foe} identification methods are widely used by militaries to authenticate their airplanes in order to detect a rival airplane, however, as can be seen in Figure \ref{fig:problems} \cite{Ruiz:2018:IRD:3236498.3214283}, these methods are not effective for drones due to the fact that a friend and a foe drone can be located too near to each other at the same altitude and GPS location. Therefore, even if the GPS location of a friend is known, it cannot be used to distinguish between friend and foe drones that are located a few centimeters from each other given the known average 4.5 meter GPS error under open skies  \cite{GPS-Accuracy}. A recent study \cite{Ruiz:2018:IRD:3236498.3214283} suggested a way of identifying each drone by matching the motion detected through  their inertial sensors and from an external camera . However, this method requires additional hardware that can obtain data from sensors and broadcast it to a drone detection mechanism using a transmitter. This method can be effective for military purposes \cite{Ruiz:2018:IRD:3236498.3214283}, however it cannot be used to authenticate/identify most commercial drones, since they do not support broadcasting their sensors' data, and out-of-band solutions to authenticate drones remain a gap.

The scientific gap in this area can be exploited for various purposes by malicious entities, particularly in areas in which drone flight is permitted. Due to the inability to distinguish between a friend and foe, in the domain of physical attacks, for example, terrorists can fly a drone equipped with a bomb in order to assassinate world leaders \cite{venezuela} or launch a multiple casualty incident \cite{ISIS,hamas}.  Other domains, including smuggling and cyber-attacks, are also affected by this scientific gap.

Considering the fact that drones do not support the functionality to authenticate and identify themselves, the ability to detect smuggling attempts and cyber-attack remains a challenge \cite{nassi2019xerox,guri2017led}. \textbf{Given  a drone that is being used to launch a cyber-attack, how can we identify this drone if its operator disguised the drone as an aerial pizza delivery drone?} We demonstrate a new type of cyber warfare against smart cities that triggers watering via a cellular smart irrigation system while disguised as a Domino's Pizza delivery drone.

In addition, some organizations deploy drones on their premises in order to transport  packages between buildings that are located some distance from one other. This can easily be exploited by attackers to spy on the organization without being detected using the same type of drone that is used by the organization.

\begin{table*}[htbp]
  \centering
  \caption{Anti Drone methods Comparison}
  \resizebox{1.0\linewidth}{!}{%
\begin{tabular}{c|c|c|c|c|c|c|c|c|}
\cline{2-9}
 & Methods & \multicolumn{3}{c|}{Effectiveness} & \multicolumn{4}{c|}{Results \& Countermeasures} \\ \cline{2-9} 
 & Type & \begin{tabular}[c]{@{}c@{}}Wi-Fi FPV\\ Drones\end{tabular} & \begin{tabular}[c]{@{}c@{}}Proprietary FPV\\ Drone\end{tabular} & \begin{tabular}[c]{@{}c@{}}LOS\\ Independent\end{tabular} & \begin{tabular}[c]{@{}c@{}}Denial of\\ Service\end{tabular} & \begin{tabular}[c]{@{}c@{}}Aerial\\ Hijacking\end{tabular} & \begin{tabular}[c]{@{}c@{}}Violating\\ Confidentiality\end{tabular} & Countermeasure \\ \hline
\multicolumn{1}{|c|}{\multirow{5}{*}{Protocol}} & Deauthentication \cite{SkyJack, Knocking-my-neighbors-kids-cruddy-drone-offline} & \Checkmark &  & \Checkmark & \Checkmark & \Checkmark &  & \multirow{3}{*}{Using WPA} \\ \cline{2-8}
\multicolumn{1}{|c|}{} & \begin{tabular}[c]{@{}c@{}}Flooding NIC \cite{Deligne2012}\end{tabular} & \Checkmark &  & \Checkmark & \Checkmark &  &  &  \\ \cline{2-8}
\multicolumn{1}{|c|}{} & \begin{tabular}[c]{@{}c@{}}Deleting Stored\\ Videos \cite{Knocking-my-neighbors-kids-cruddy-drone-offline}\end{tabular} & \Checkmark &  & \Checkmark &  &  &  &  \\ \cline{2-9}
\multicolumn{1}{|c|}{} & \begin{tabular}[c]{@{}c@{}}Replay\\ Attacks \cite{Hacking-A-Professional-Drone,highnam2016uncrewed}\end{tabular} &  & \Checkmark & \Checkmark &  & \Checkmark &  & \begin{tabular}[c]{@{}c@{}}Operator's\\ Authentication \cite{shoufan2017continuous}\end{tabular}  \\ \cline{2-9} 
\multicolumn{1}{|c|}{} & \begin{tabular}[c]{@{}c@{}}Detecting Captured\\ POI \cite{nassi2018game}\end{tabular} & \Checkmark &  &  &  &  & \Checkmark & \begin{tabular}[c]{@{}c@{}}Disabling Video\\ Compression\end{tabular} \\ \hline
\multicolumn{1}{|c|}{\multirow{4}{*}{\begin{tabular}[c]{@{}c@{}}Spoofing\\ Sensors\end{tabular}}} & Camera \cite{davidson2016controlling} & \Checkmark & \Checkmark &  &  & Shifting &  &  \\ \cline{2-9} 
\multicolumn{1}{|c|}{} & \begin{tabular}[c]{@{}c@{}} GPS\\ \cite{Drones-Hijacking, kerns2014unmanned, Knocking-my-neighbors-kids-cruddy-drone-offline}\end{tabular} & \Checkmark & \Checkmark & \Checkmark & \begin{tabular}[c]{@{}c@{}}Force\\ Landing\end{tabular} & \begin{tabular}[c]{@{}c@{}}During RTH\\ Mission\end{tabular} &  & \begin{tabular}[c]{@{}c@{}}Anti GPS\\Spoofing \cite{feng2017efficient,feng2018efficient}\end{tabular}  \\ \cline{2-9} 
\multicolumn{1}{|c|}{} & \begin{tabular}[c]{@{}c@{}}Motion\\ Sensors \cite{son2015rocking}\end{tabular} & \Checkmark & \Checkmark & \Checkmark & Landing &  &  & \begin{tabular}[c]{@{}c@{}}Software Based\\ Solution \cite{choi2018detecting}\end{tabular} \\ \cline{2-9} 
\multicolumn{1}{|c|}{} & Magnometer \cite{Knocking-my-neighbors-kids-cruddy-drone-offline} & \Checkmark & \Checkmark & \Checkmark & \begin{tabular}[c]{@{}c@{}}Force\\ Calibration\end{tabular} &  &  &  \\ \hline
\multicolumn{1}{|c|}{\multirow{2}{*}{\begin{tabular}[c]{@{}c@{}}Compromised\\ Component\end{tabular}}} & \begin{tabular}[c]{@{}c@{}}Fake\\ Propeller \cite{206156}\end{tabular} & \Checkmark & \Checkmark & \Checkmark & Crashing &  &  & \begin{tabular}[c]{@{}c@{}}Parachute \\ \cite{parazero, MARS-Parachutes, fruitychutes}\end{tabular} \\ \cline{2-9} 
\multicolumn{1}{|c|}{} & \begin{tabular}[c]{@{}c@{}}Compromised\\ Firmware \cite{Maldrone,Drones-Hijacking}\end{tabular} & \Checkmark & \Checkmark & \Checkmark &  & \begin{tabular}[c]{@{}c@{}}Using a\\ Backdoor\end{tabular} &  & \begin{tabular}[c]{@{}c@{}}Control Flow\\ Approach \cite{pike2016trackos}\end{tabular} \\ \hline
\multicolumn{1}{|c|}{\multirow{2}{*}{Jammers}} & GPS \cite{Knocking-my-neighbors-kids-cruddy-drone-offline} & \Checkmark & \Checkmark & \Checkmark & \begin{tabular}[c]{@{}c@{}}Drifting, Loss\\ of Control\end{tabular} &  &  &  \\ \cline{2-9} 
\multicolumn{1}{|c|}{} & FPV Channel \cite{Drones-Hijacking} & \Checkmark & \Checkmark & \Checkmark & \begin{tabular}[c]{@{}c@{}}Disabling\\ FPV\end{tabular} &  &  &  \\ \hline
\multicolumn{1}{|c|}{\multirow{5}{*}{\begin{tabular}[c]{@{}c@{}}Physical\\ Attacks\end{tabular}}} & Nets & \Checkmark & \Checkmark &  & \begin{tabular}[c]{@{}c@{}}Landing,\\ Crashing\end{tabular} &  &  &  \\ \cline{2-9} 
\multicolumn{1}{|c|}{} & Bullets & \Checkmark & \Checkmark &  & Crashing &  &  & \begin{tabular}[c]{@{}c@{}}Parachute \\ \cite{parazero, MARS-Parachutes, fruitychutes}\end{tabular} \\ \cline{2-9} 
\multicolumn{1}{|c|}{} & Lasers & \Checkmark & \Checkmark &  & Crashing &  &  & \begin{tabular}[c]{@{}c@{}}Mirrors, Smoke\end{tabular} \\ \cline{2-9} 
\multicolumn{1}{|c|}{} & Missiles & \Checkmark & \Checkmark &  & Crashing &  &  &  \\ \cline{2-9} 
\multicolumn{1}{|c|}{} & \begin{tabular}[c]{@{}c@{}}Predator Birds\end{tabular} & \Checkmark & \Checkmark &  & Landing &  &  &  \\ \hline
\end{tabular}
}
\end{table*}%

\section{Attacks Against Drones \& Countermeasures}
\label{sec:Disabling}

In this section, we review methods used to attack drones. Attacks against drones can result in the following:
\begin{itemize}
    \item Hijacking - The attacker gains complete control of the drone.
    \item Denial-of-Service - The attacker causes the drone to one of the following results: crashing, landing, drifting, disabling the operator the video link. 
    \item Violating confidentiality - The attacker can determine whether a specific point of interest is being video streamed by the drone.
\end{itemize}
In addition, we also review countermeasures methods against the attacks.

\subsection{Protocol-Based Attacks \& Countermeasures}
In this subsection, we review attacks that exploit protocol vulnerabilities. Older drone generations that were manufactured by Parrot were based on Wi-Fi FPV that did not require any authentication to join the network (open access points). Several studies \cite{altawy2017security, SkyJack, Deligne2012} applied traditional methods  against open access points and presented two DOS attacks against these drones by (1) applying a deauthentication attack \cite{SkyJack, Knocking-my-neighbors-kids-cruddy-drone-offline}, and (2) flooding the drone's NIC \cite{Deligne2012}. Performing a hijacking attack after the deauthentication attack stage was also suggested by \cite{SkyJack}. Another study \cite{Knocking-my-neighbors-kids-cruddy-drone-offline} found that the FTP folder that stores images and videos captured by Parrot drones does not require any authentication and deleted files from the FTP folder. However, the abovementioned methods can only be applied to older Parrot drones models as new Parrot drones support WPA-2 with authentication; therefore, as long as the joining  password has not been leaked, the abovementioned attacks cannot be applied  on newer Parrot drones. Two studies demonstrated  methods for hijacking a drone using a replay attack that was applied from a malicious GCS against weak uplinks of FPV channels (Figure \ref{fig:FPV}).  A novel study \cite{Hacking-A-Professional-Drone} presented techniques used to hijack a \$30k drone used by police departments by exploiting the XBee 868LP protocol using replaying maneuvering commands that are sent over 868 MHz from the GCD  to the drone. Another study \cite{highnam2016uncrewed} showed that amateur drones whose uplinks are based on the MAVLink protocol and can be found on amateur drones  (e.g., 3DR IRIS+, Erle-Copter) can also be hijacked using a replay attack. However, a recent \cite{shoufan2017continuous} study showed that malicious replay attacks can be detected by authenticating the drone's operator via measurements obtained from motion sensors using machine learning algorithms.

\subsection{Sensor-Based Attacks \& Countermeasures}
As described in Section \ref{Sec:background}, drones contain motion sensors, obstacle avoidance sensors, cameras, and many other sensors that are important for real-time maneuvering. In this subsection, we review sensor-based attacks, which are also known as spoofing attacks. The MEMS gyroscope and accelerometers are sensitive to ultrasound at their resonance frequency, and attacks targeting this vulnerability have been demonstrated in many studies \cite{trippel2017walnut,farshteindiker2016phone} and was demonstrated in an article from USENIX 2015 by \cite{son2015rocking} where the authors spoofed a drone's gyroscope output to maximum values,  forcing the drone to land. However, this type of attack requires powerful speakers and is very limited in its range, since sound deteriorates with distance. A recent paper from CCS 2018 \cite{choi2018detecting} proposed a software solution for the acoustic attack  that was presented at USENIX 2015 \cite{son2015rocking}. The authors implemented a software-based solution for control invariant checking and demonstrated that sensor spoofing attacks that result in anomalous measurements obtained from a drone's sensors can be detected by comparing a measurement that was obtained from a sensor to a measurement that was predicted by their framework. Another study \cite{davidson2016controlling} demonstrated a method of hijacking a drone by spoofing its downward camera and influencing the stabilizing algorithm (which is based on detecting movement changes from a video stream) by directing a laser and projector to the surface of a flying drone. A few studies \cite{Drones-Hijacking, kerns2014unmanned, Knocking-my-neighbors-kids-cruddy-drone-offline, vervisch2017influence, he2018friendly} presented a method for hijacking and disabling a drone using GPS spoofing (1) of no-fly zones, (2) during autonomous navigation to a target.  In order to detect GPS spoofing attacks, various software \cite{feng2017efficient,feng2018efficient} and hardware-based solutions for drones were adopted from prior knowledge that exists in this area . Another study \cite{Knocking-my-neighbors-kids-cruddy-drone-offline} found that the presence of a magnetic field near the DJI Phantom 3 always necessitated recalibration of the drone's compass prior to takeoff.

\subsection{Compromised Component \& Countermeasures}
Recently, a new type of attack targeted at drones was suggested by \cite{206156,Maldrone,Drones-Hijacking} based on compromising their hardware/software. A recent study \cite{206156} demonstrated a supply chain attack against drone hardware using a propeller that is visually identical to a genuine propeller but crashes the drone upon takeoff.  Two other studies suggested attacks against drone software. In \cite{Maldrone}, the researcher installed malware on a drone's firmware and used it to open a backdoor. In \cite{Drones-Hijacking}, the researcher reverse engineered DJI's SDK  and escalated  its permissions in order to create a C\&C application for smartphones which is used to control the drone. Compromising a drone's software can be accomplished with a supply chain attack or via the Internet by compromising a drone's Internet connected GCD (e.g., a smartphone) with malware or a fake application. A recent study \cite{pike2016trackos} suggested a control flow integrity approach to detect attacks against software. They implemented their approach on the ArduPilot OS and showed that it was capable of detecting attacks such as buffer overflow attacks and illegal executions of functions.

\subsection{Jammers}
Several studies have suggested disrupting incoming/outgoing communication using jammers. One study showed that applying GPS jamming \cite{Knocking-my-neighbors-kids-cruddy-drone-offline} to drones results in drifting and difficulty controlling the drone, and prevents the return to home functionality from working. Another study applied radio jamming \cite{Drones-Hijacking} against a video link channel and showed that the FPV functionality was disabled in the GCD , preventing the operator from maneuvering the drone with no LOS. Jammers are one of the most commonly used products for disabling a drone on the market. Some commercial anti-drone jammers are directional RF transmitters in the form of mobile shooting guns that apply jamming to GPS signals and ISM bands known to be used by drones \cite{DroneDefender, AUDS1, DroneJammerGun, DroneGun, SKYNET, Scrambler, Dronebuster}. Other jammers are stationary devices \cite{Dynopis, Xpeller, MCHORIZON, DroneDefeater}. A jammer's ability relies on the strength of its radio transmitter, however the best jammer on the market is effective at ranges of up to two kilometers.

\subsection{Physical Attacks \& Countermeasures}
A physical attack is the most common means of disabling drones used by the industrial sector. Several companies utilize nets to disable and crash drones. Such nets are connected to drones that swoop and swag malicious drones or fire a shot from the air (using another drone) \cite{OneTouchInterceptor,DroneCatcher,DroneHunter,RoboticFalconry,SparrowHawk}, or the ground (using a gun) \cite{NetGun,NetGunX1,SkyWall,ShotgunShellNets}. The net stops the propellers from turning and causes the drone to fall and crash to the ground. One company \cite{DroneHuntingEagles} sells predator birds (eagles and falcons) that have been trained to detect, capture, and land drones. Other companies sell vehicles equipped with laser guns and cannons that irradiate a directed high energy laser beam that causes a drone to burn in the air and fall to the ground \cite{IronBeam,LaserEffector}. We also note that an army has used a \$3 million Patriot missile to shoot down a drone \cite{patriot}. Since physical attacks can cause the drone to crash and in so doing harm people (in populated areas that allow drone flight) and cause damage to facilities and the drone itself, several companies sell a dedicated parachute for drones (e.g., DJI Inspire 1 and 2, DJI Phantom 2-4, DJI Matrice) \cite{parazero, MARS-Parachutes, fruitychutes}. Some of the parachutes contain an automatic trigger system that is designed to open the parachute if the drone falls , while others are manually opened by the drone operator (based on the RF controller). Some parachutes activate an audio buzzer to warn bystanders to move out of harm's way.

\section{Future Research Directions}
\label{sec:future-research-directions}

In this section, we suggest future research directions in areas in which a scientific gap exists. We include suggestions that can be applied  in order to detect attacks caused by malicious operators who will not follow regulations and laws.

\subsection{Suggestions for Authenticating Drones and Operators}

Methods for authenticating drones based on a white-listing approach must be introduced in order to solve the identification problem (i.e., detecting a specific drone among similar drones).  For example, one interesting method that can be used for this purpose as an out-of-band solution is installing  a microcontroller on a group of white-listed drones.  In this case, the microcontroller will serve as a transmitter for authenticating the drone by modulating an RSA token using visual cues (e.g., using an LED strip); the visual cues will be captured by a video camera connected to a computer that will analyze the frames in the video stream and interpret the series of visual modulations to an RSA token . Based on the result, it  will decide to authenticate the drone or not. We also hypothesize whether drone operators can be authenticated based on their flying skills captured via a third party static radio receiver by analyzing a drone’s radio emissions. While authenticating a drone operator has already been suggested using an on board approach  (which we do not consider a third party approach), we propose using the method introduced in \cite{zeng2016wiwho} for authenticating smartphone users in indoor environments via a laptop’s network interface card by analyzing the channel state information of the Wi-Fi protocol .
are based on replay attacks by analyzing the received signal strength indication (RSSI) of a radio command that was sent from another GCS (i.e., if the difference in the received signal strength indication between two consecutive commands is over a predefined threshold, the received command will be ignored).



\subsection{Suggestions for Dealing with Cyber-Attacks on Drones}

While detecting a cyber-attack against drones is difficult, trying to deal with the repercussions of cyber-attacks is much harder. For example, even if a sensor-based attack has been detected, securing the drone or returning it home safely is a complex problem if the drone’s system  cannot rely on the measurements from the integrated sensors (e.g., GPS, gyroscope). In order to deal with this problem, we hypothesize how accurate will be a software-based mechanism that stores the history of the series of maneuvering commands (from the time the drone has takeoff until a sensor-based attack has been identified,  e.g., using the technique suggest by \cite{choi2018detecting}). When an attack (e.g., a hybrid GPS spoofing and FPV jamming attack) that prevents the drone from flying according to the commands sent from its operator (due to FPV jamming) and automatically returning home (due to GPS spoofing) has been detected, the drone can return to its takeoff location by traversing  its series of maneuvering commands from the last command (when the attack was identified) to the first command (takeoff) and performing the opposite of each of the commands.

\subsection{Suggestions for Determining a Drone’s Intention}
Other mechanisms for detecting the purpose of a drone based on its flight behavior must also be introduced. For example, we hypothesize  whether the cellular hijacking that was demonstrated in our video  can be detected by analyzing the radio activity of known cellular bands using a spectrum analyzer and intersecting  the origin of the cellular transmitter with the location of a detected drone.

\subsection{Open Questions}
One important question that requires a solution is how to detect and locate the operator of a malicious drone given the fact that such malicious operators will not follow regulations, such as installing a form of identification on a drone or registering its unique identifier in a national database, even if such regulations are instituted. In addition, we hypothesize  whether there is a method for creating a unique signature for each drone that cannot be changed or copied (for example, a dedicated acoustic signature resulting from the manufacturing process of a drone’s rotor).

\bibliographystyle{IEEEtran}
\bibliography{IEEEabrv,main}

\begin{thebibliography}{100}
\providecommand{\url}[1]{#1}
\csname url@samestyle\endcsname
\providecommand{\newblock}{\relax}
\providecommand{\bibinfo}[2]{#2}
\providecommand{\BIBentrySTDinterwordspacing}{\spaceskip=0pt\relax}
\providecommand{\BIBentryALTinterwordstretchfactor}{4}
\providecommand{\BIBentryALTinterwordspacing}{\spaceskip=\fontdimen2\font plus
\BIBentryALTinterwordstretchfactor\fontdimen3\font minus
  \fontdimen4\font\relax}
\providecommand{\BIBforeignlanguage}[2]{{%
\expandafter\ifx\csname l@#1\endcsname\relax
\typeout{** WARNING: IEEEtran.bst: No hyphenation pattern has been}%
\typeout{** loaded for the language `#1'. Using the pattern for}%
\typeout{** the default language instead.}%
\else
\language=\csname l@#1\endcsname
\fi
#2}}
\providecommand{\BIBdecl}{\relax}
\BIBdecl

\bibitem{terrorism-by-joystick-1}
P.~Gazette, ``Terrorism by joystick,''
  \url{http://www.post-gazette.com/opinion/2018/08/07/Terrorism-by-joystick/stories/201808070022}.

\bibitem{terrorism-by-joystick-2}
W.~Post, ``Drone attacks are essentially terrorism by joystick,''
  \url{https://www.washingtonpost.com/opinions/drone-attacks-are-essentially-terrorism-by-joystick/2018/08/05/f93ec18a-98d5-11e8-843b-36e177f3081c_story.html?noredirect=on&utm_term=.faccb2a27e26}.

\bibitem{terrorism-by-joystick-3}
S.~C.~M. Post, ``Analysis: with drone attacks, the era of joystick terrorism
  appears to have arrived,''
  \url{https://www.scmp.com/news/world/article/2158380/analysis-drone-attacks-prove-era-joystick-terrorism-has-arrived-and-world}.

\bibitem{PIZZA-DELIVERY-BY-DRONE-LAUNCHED-BY-DOMINOS}
Newsweek, ``Pizza delivery by drone launched by domino's,''
  \url{http://www.newsweek.com/pizza-delivery-drone-dominos-493371}.

\bibitem{Amazon}
B.~Insider, ``Amazon and ups are betting big on drone delivery,''
  \url{http://www.businessinsider.com/amazon-and-ups-are-betting-big-on-drone-delivery-2018-3}.

\bibitem{CNN-Just-Got-Approved-to-Fly-Drones-Over-Crowds-of-People}
Fortune, ``Cnn just got approved to fly drones over crowds of people,''
  \url{http://fortune.com/2017/10/18/cnn-drones-faa-crowds/}, 2017.

\bibitem{cheating}
N.~Y. Post, ``Husband uses drone to catch cheating wife,''
  \url{https://nypost.com/2016/11/16/husband-uses-drone-to-catch-cheating-wife/},
  2016.

\bibitem{26-floor}
kiro7, ``Woman terrified by drone outside her window,''
  \url{http://www.kiro7.com/news/woman-terrified-drone-outside-her-window/81721261},
  2014.

\bibitem{Woman-grabs-gun-shoots-nosy-neighbour-s-drone}
D.~Mail, ``Woman grabs gun shoots nosy neighbour's drone,''
  \url{http://www.dailymail.co.uk/news/article-4283486/Woman-grabs-gun-shoots-nosy-neighbour-s-drone.html}.

\bibitem{Virginia-Woman-Shoots-Down-Drone-Near-Actor-Robert-Duvalls-Home}
N.~Washington, ``Virginia woman shoots down drone near actor robert duvalls
  home,''
  \url{http://www.nbcwashington.com/news/local/Virginia-Woman-Shoots-Down-Drone-Near-Actor-Robert-Duvalls-Home-391423411.html}.

\bibitem{not-my-backyard-man-arrested-after-shooting-drone-down}
N.~News, ``Kentucky man arrested after shooting down neighbor's drone,''
  \url{http://www.nbcnews.com/news/us-news/not-my-backyard-man-arrested-after-shooting-drone-down-n402271}.

\bibitem{Big-rise-in-drone-jail-smuggling-incidents}
BBC, ``Big rise in drone jail smuggling incidents,''
  \url{http://www.bbc.com/news/uk-35641453}.

\bibitem{Two-plead-guilty-in-border-drug-smuggling-by-drone}
L.~A. Times, ``Two plead guilty in border drug smuggling by drone,''
  \url{http://www.latimes.com/local/california/la-me-drone-drugs-20150813-story.html}.

\bibitem{smuggling-iphones}
D.~Trends, ``Smugglers used aerial drones to sneak \$80 million in iphones into
  china,''
  \url{https://www.digitaltrends.com/mobile/iphone-smugglers-aerial-drones-hong-kong-china/}.

\bibitem{smuggling-tobaco}
T.~Drive, ``Tobacco-smuggling drone found by ukraine border patrol reveals
  region's black market,''
  \url{http://www.thedrive.com/tech/23447/tobacco-smuggling-drone-found-by-ukraine-border-patrol-reveals-regions-black-market/}.

\bibitem{venezuela}
N.~Y. Times, ``Venezuelan president targeted by drone attack, officials say,''
  \url{https://www.nytimes.com/2018/08/04/world/americas/venezuelan-president-targeted-in-attack-attempt-minister-says.html}.

\bibitem{drone-marketshare}
B.~Insider, ``Commercial unmanned aerial vehicle (uav) market analysis,''
  \url{http://www.businessinsider.com/commercial-uav-market-analysis-2017-8},
  2017.

\bibitem{drone-marketshare-2}
------, ``Drone market shows positive outlook with strong industry growth and
  trends,''
  \url{http://www.businessinsider.com/drone-industry-analysis-market-trends-growth-forecasts-2017-7},
  2017.

\bibitem{President-Trump-Moves-to-Fill-America's-Skies-with-Drones}
Wired, ``President trump moves to fill america's skies with drones,''
  \url{https://www.wired.com/story/faa-trump-drones-regulations/}, 2017.

\bibitem{anti-drone-market}
``The global anti-drone market size is anticipated to reach usd 1.85 billion by
  2024,''
  \url{https://www.prnewswire.com/news-releases/the-global-anti-drone-market-size-is-anticipated-to-reach-usd-1-85-billion-by-2024--300673188.html},
  2018.

\bibitem{altawy2017security}
R.~Altawy and A.~M. Youssef, ``Security, privacy, and safety aspects of
  civilian drones: A survey,'' \emph{ACM Transactions on Cyber-Physical
  Systems}, vol.~1, no.~2, p.~7, 2017.

\bibitem{guvenc2018detection}
I.~Guvenc, F.~Koohifar, S.~Singh, M.~L. Sichitiu, and D.~Matolak, ``Detection,
  tracking, and interdiction for amateur drones,'' \emph{IEEE Communications
  Magazine}, vol.~56, no.~4, pp. 75--81, 2018.

\bibitem{guvencc2017detection}
{\.I}.~G{\"u}ven{\c{c}}, O.~Ozdemir, Y.~Yapici, H.~Mehrpouyan, and D.~Matolak,
  ``Detection, localization, and tracking of unauthorized uas and jammers,'' in
  \emph{Digital Avionics Systems Conference (DASC), 2017 IEEE/AIAA 36th}.\hskip
  1em plus 0.5em minus 0.4em\relax IEEE, 2017, pp. 1--10.

\bibitem{game-of-drones}
F.~Brown, ``Game of drones,'' \emph{DefCon 25}, 2017.

\bibitem{sturdivant2017systems}
R.~L. Sturdivant and E.~K. Chong, ``Systems engineering baseline concept of a
  multispectral drone detection solution for airports,'' \emph{IEEE Access},
  vol.~5, pp. 7123--7138, 2017.

\bibitem{shi2018anti}
X.~Shi, C.~Yang, W.~Xie, C.~Liang, Z.~Shi, and J.~Chen, ``Anti-drone system
  with multiple surveillance technologies: Architecture, implementation, and
  challenges,'' \emph{IEEE Communications Magazine}, vol.~56, no.~4, pp.
  68--74, 2018.

\bibitem{michel2018counter}
A.~H. Michel, \emph{Counter-drone systems}.\hskip 1em plus 0.5em minus
  0.4em\relax Center for the Study of the Drone at Bard College, 2018.

\bibitem{hoversurf}
H.~Surf, ``hoversurf - evtol,'' \url{https://www.hoversurf.com/}.

\bibitem{SureFly}
B.~Insider, ``Surefly,'' \url{https://www.youtube.com/watch?v=w8W3aBgoRcU}.

\bibitem{ehang}
ehang, ``ehang184,'' \url{http://www.ehang.com/ehang184/}.

\bibitem{WiFi-FPV-vs-5.8GHz-FPV-vs-2.4GHz-FPV-Ultimate-Guide}
rcdronearena, ``Wifi fpv vs 5.8ghz fpv vs 2.4ghz fpv: Ultimate guide,''
  \url{http://www.rcdronearena.com/2016/03/15/wifi-fpv-vs-5-8ghz-fpv-vs-2-4ghz-fpv-explained/}.

\bibitem{WiFi-FPV-vs-5.8GHz-FPV-vs-2.4GHz-FPV}
B.~Quadcopter, ``Wifi fpv vs 5.8ghz fpv vs 2.4ghz fpv,''
  \url{https://www.best-quadcopter.com/versus-zone/2016/04/wifi-fpv-vs-5-8ghz-fpv-vs-2-4ghz-fpv/}.

\bibitem{Wifi-FPV-Drones}
auselectronicsdirect, ``Wifi fpv drones,''
  \url{https://www.auselectronicsdirect.com.au/drones/fpv-drone/wifi-fpv-drones/}.

\bibitem{8-smartphone-controlled-drones}
androidauthority, ``8 fun drones you can control with your smartphone,''
  \url{https://www.androidauthority.com/best-smartphone-controlled-drones-744632/}.

\bibitem{8-DRONES-THAN-CAN-BE-CONTROLLED-BY-A-SMARTPHONE}
dronesglobe, ``8 drones than can be controlled by a smartphone (fully or
  partially),'' \url{http://www.dronesglobe.com/guide/smartphone-drones/}.

\bibitem{Skydio}
Skydio, ``Skydio r1,'' \url{https://www.skydio.com/}.

\bibitem{FAA-committee-says-small-drones-should-be-allowed-to-fly-over-cities-and-crowds}
T.~Verge, ``Faa committee says small drones should be allowed to fly over
  cities and crowds,''
  \url{https://www.theverge.com/2016/4/6/11362900/faa-small-drone-recommendation-cities-urban-areas},
  2016.

\bibitem{UPS-FEDEX}
globalresearch, ``Delivery drones and driverless vehicles: Ups, fedex investing
  in drones,''
  \url{https://www.globalresearch.ca/delivery-drones-and-driverless-vehicles-ups-fedex-investing-in-drones/5614041}.

\bibitem{dji-agriculture}
DJI, ``Agrasmg-1,'' \url{https://www.dji.com/mg-1}.

\bibitem{IDF-DJI}
YNET, ``Idf infantry company commanders to receive drones,''
  \url{https://www.ynetnews.com/articles/0,7340,L-4960749,00.html}.

\bibitem{NZ-DJI}
nzherald, ``Nzdf has no plans to ground drones banned by us military allies
  over cyber-safety fears,''
  \url{https://www.nzherald.co.nz/nz/news/article.cfm?c_id=1&objectid=12005158}.

\bibitem{drones-applications}
G.~World, ``10 major application areas of drone,''
  \url{https://www.geospatialworld.net/blogs/10-major-application-areas-of-drone/}.

\bibitem{top-12-non-military-uses-for-drone}
A.~D. Craze, ``Top 12 non military uses for drone,''
  \url{https://www.airdronecraze.com/drones-action-top-12-non-military-uses/}.

\bibitem{top-38-non-military-uses-for-drone}
C.~Insights, ``38 ways drones will impact society: From fighting war to
  forecasting weather, uavs change everything,''
  \url{https://www.cbinsights.com/research/drone-impact-society-uav/}.

\bibitem{Autonomous-Killer-Drones}
YouTube, ``Autonomous killer drones,''
  \url{https://www.youtube.com/watch?v=DK6IGG5zRU8}.

\bibitem{Burglarss}
telegraph, ``Burglars use drone helicopters to target homes,''
  \url{https://www.telegraph.co.uk/news/uknews/crime/11613568/Burglars-use-drone-helicopters-to-identify-targe-homes.html}.

\bibitem{karanam20173d}
C.~R. Karanam and Y.~Mostofi, ``3d through-wall imaging with unmanned aerial
  vehicles using wifi,'' in \emph{Proceedings of the 16th ACM/IEEE
  International Conference on Information Processing in Sensor Networks}.\hskip
  1em plus 0.5em minus 0.4em\relax ACM, 2017, pp. 131--142.

\bibitem{imsi-catcher}
E.~Group, ``Milipol 2017: Eca group unveils its signal intelligence solution
  for mounting embedded on its aerial,''
  \url{https://www.ecagroup.com/en/business/milipol-2017-eca-group-unveils-its-signal-intelligence-solution-mounting-embedded-its}.

\bibitem{Snoopy}
\BIBentryALTinterwordspacing
G.~Wilkinson, ``The machines that betrayed their masters,'' 2014. [Online].
  Available: \url{https://www.youtube.com/watch?v=GvrB6S_O0BE}
\BIBentrySTDinterwordspacing

\bibitem{Japan-PM}
Spiegel, ``Man arrested for landing 'radioactive' drone on japanese prime
  minister's roof,''
  \url{https://www.independent.co.uk/news/world/asia/man-arrested-for-landing-radioactive-drone-on-japanese-prime-ministers-roof-10203517.html}.

\bibitem{UAV-Related-Incidents}
Wikipedia, ``Uav related incidents,''
  \url{https://en.wikipedia.org/wiki/List_of_UAV-related_incidents}.

\bibitem{Gatwick}
------, ``Gatwick airport drone incident,''
  \url{https://en.wikipedia.org/wiki/Gatwick_Airport_drone_incident}.

\bibitem{chesaux2014wireless}
J.~Chesaux, ``Wireless access point spoofing and mobile devices geolocation
  using swarms of flying robots,'' \emph{Master optional semester project,
  Spring}, 2014.

\bibitem{nassi2019xerox}
B.~Nassi, A.~Shamir, and Y.~Elovici, ``Xerox day vulnerability,'' \emph{IEEE
  Transactions on Information Forensics and Security}, vol.~14, no.~2, pp.
  415--430, 2019.

\bibitem{guri2017led}
M.~Guri, B.~Zadov, and Y.~Elovici, ``Led-it-go: Leaking (a lot of) data from
  air-gapped computers via the (small) hard drive led,'' in \emph{International
  Conference on Detection of Intrusions and Malware, and Vulnerability
  Assessment}.\hskip 1em plus 0.5em minus 0.4em\relax Springer, 2017, pp.
  161--184.

\bibitem{attack-against-russia}
T.~Drive, ``Russia offers new details about syrian mass drone attack, now
  implies ukrainian connection,''
  \url{http://www.thedrive.com/the-war-zone/17595/russia-offers-new-details-about-syrian-mass-drone-attack-now-implies-ukrainian-connection}.

\bibitem{ronen2017iot}
E.~Ronen, A.~Shamir, A.-O. Weingarten, and C.~O'Flynn, ``Iot goes nuclear:
  Creating a zigbee chain reaction,'' in \emph{Security and Privacy (SP), 2017
  IEEE Symposium on}.\hskip 1em plus 0.5em minus 0.4em\relax IEEE, 2017, pp.
  195--212.

\bibitem{197130}
\BIBentryALTinterwordspacing
K.~Fawaz, K.-H. Kim, and K.~G. Shin, ``Protecting privacy of {BLE} device
  users,'' in \emph{25th {USENIX} Security Symposium ({USENIX} Security
  16)}.\hskip 1em plus 0.5em minus 0.4em\relax Austin, TX: {USENIX}
  Association, 2016, pp. 1205--1221. [Online]. Available:
  \url{https://www.usenix.org/conference/usenixsecurity16/technical-sessions/presentation/fawaz}
\BIBentrySTDinterwordspacing

\bibitem{laser-microphone}
Wikipedia, ``Laser microphone,''
  \url{https://en.wikipedia.org/wiki/Laser_microphone}.

\bibitem{A-Drone-Too-Small-for-Radar-to-Detect-Rattles-the-White-House}
N.-Y. Times, ``A drone, too small for radar to detect, rattles the white
  house,'' \url{https://www.nytimes.com/2015/01/27/us/white-house-drone.html}.

\bibitem{greenpeace}
Reuters, ``Greenpeace crashes superman-shaped drone into french nuclear
  plant,''
  \url{https://www.reuters.com/article/us-france-nuclear-greenpeace/greenpeace-crashes-superman-shaped-drone-into-french-nuclear-plant-idUSKBN1JT1JM}.

\bibitem{ISIS}
CNN, ``Fears over isis use of bomb-carrying drones,''
  \url{https://edition.cnn.com/videos/world/2018/06/18/isis-bomb-carrying-drones-fears-todd-pkg.cnn}.

\bibitem{hamas}
T.~Y. World, ``New threat: Hamas using drones to drop explosives on israeli
  towns near gaza,''
  \url{https://www.theyeshivaworld.com/news/israel-news/1566574/new-threat-hamas-using-drones-to-drop-explosives-on-israeli-towns-near-gaza.html}.

\bibitem{German-Police}
Independent, ``German police shoot down model plane terror plot,''
  \url{http://www.spiegel.de/international/germany/german-police-suspect-remote-controlled-airplane-terror-plot-a-907756.html}.

\bibitem{attacks-against-airplanes}
Wired, ``A drone-flinging cannon proves uavs can mangle planes,''
  \url{https://www.wired.com/story/drone-plane-collision-damage-study/}.

\bibitem{toh2017cyber}
J.~Toh, M.~Hatib, O.~Porzecanski, and Y.~Elovici, ``Cyber security patrol:
  detecting fake and vulnerable wifi-enabled printers,'' in \emph{Proceedings
  of the Symposium on Applied Computing}.\hskip 1em plus 0.5em minus
  0.4em\relax ACM, 2017, pp. 535--542.

\bibitem{zhang2017dolphinattack}
G.~Zhang, C.~Yan, X.~Ji, T.~Zhang, T.~Zhang, and W.~Xu, ``Dolphinattack:
  Inaudible voice commands,'' in \emph{Proceedings of the 2017 ACM SIGSAC
  Conference on Computer and Communications Security}.\hskip 1em plus 0.5em
  minus 0.4em\relax ACM, 2017, pp. 103--117.

\bibitem{carlini2016hidden}
N.~Carlini, P.~Mishra, T.~Vaidya, Y.~Zhang, M.~Sherr, C.~Shields, D.~Wagner,
  and W.~Zhou, ``Hidden voice commands.'' in \emph{USENIX Security Symposium},
  2016, pp. 513--530.

\bibitem{vanhoef2017key}
M.~Vanhoef and F.~Piessens, ``Key reinstallation attacks: Forcing nonce reuse
  in wpa2,'' in \emph{Proceedings of the 2017 ACM SIGSAC Conference on Computer
  and Communications Security}.\hskip 1em plus 0.5em minus 0.4em\relax ACM,
  2017, pp. 1313--1328.

\bibitem{bihambreaking}
E.~Biham and L.~Neumann, ``Breaking the bluetooth pairing--fixed coordinate
  invalid curve attack.''

\bibitem{shooting-drone}
Hacked, ``Flying gun? armed drone robbery? amateur weaponized drone built
  (video),''
  \url{https://hacked.com/flying-gun-armed-drone-robbery-amateur-weaponized-drone-built-video/}.

\bibitem{drone-robber-bank}
youTube, ``First bank robbery movie with a drone - the drone job,''
  \url{https://www.youtube.com/watch?v=mUDQe8vtrLM}.

\bibitem{drone-robber-subject}
Facebook, ``Drone thief!''
  \url{https://www.facebook.com/RTvids/videos/715799642116628/}.

\bibitem{Secret-Service-Arrests-Man-After-Drone-Flies-Near-White-House}
N.-Y. Times, ``Secret service arrests man after drone flies near white house,''
  \url{https://www.nytimes.com/2015/05/15/us/white-house-drone-secret-service.html}.

\bibitem{eshel2013mobile}
T.~Eshel, ``Mobile radar optimized to detect uavs, precision guided weapons,''
  \emph{Defense Update}, 2013.

\bibitem{10GHz-radar}
F.~I. for High Frequency~Physics and R.~T. FHR, ``Detection of small drones
  with millimeter wave radar,''
  \url{https://www.fhr.fraunhofer.de/en/businessunits/security/Detection-of-small-drones-with-millimeter-wave-radar.html}.

\bibitem{drozdowicz201635}
J.~Drozdowicz, M.~Wielgo, P.~Samczynski, K.~Kulpa, J.~Krzonkalla, M.~Mordzonek,
  M.~Bryl, and Z.~Jakielaszek, ``35 ghz fmcw drone detection system,'' in
  \emph{Radar Symposium (IRS), 2016 17th International}.\hskip 1em plus 0.5em
  minus 0.4em\relax IEEE, 2016, pp. 1--4.

\bibitem{de2016flight}
M.~U. de~Haag, C.~G. Bartone, and M.~S. Braasch, ``Flight-test evaluation of
  small form-factor lidar and radar sensors for suas detect-and-avoid
  applications,'' in \emph{Digital Avionics Systems Conference (DASC), 2016
  IEEE/AIAA 35th}.\hskip 1em plus 0.5em minus 0.4em\relax IEEE, 2016, pp.
  1--11.

\bibitem{molchanov2014classification}
P.~Molchanov, R.~I. Harmanny, J.~J. de~Wit, K.~Egiazarian, and J.~Astola,
  ``Classification of small uavs and birds by micro-doppler signatures,''
  \emph{International Journal of Microwave and Wireless Technologies}, vol.~6,
  no. 3-4, pp. 435--444, 2014.

\bibitem{liu2017digital}
Y.~Liu, X.~Wan, H.~Tang, J.~Yi, Y.~Cheng, and X.~Zhang, ``Digital television
  based passive bistatic radar system for drone detection,'' in \emph{Radar
  Conference (RadarConf), 2017 IEEE}.\hskip 1em plus 0.5em minus 0.4em\relax
  IEEE, 2017, pp. 1493--1497.

\bibitem{hoffmann2016micro}
F.~Hoffmann, M.~Ritchie, F.~Fioranelli, A.~Charlish, and H.~Griffiths,
  ``Micro-doppler based detection and tracking of uavs with multistatic
  radar,'' in \emph{Radar Conference (RadarConf), 2016 IEEE}.\hskip 1em plus
  0.5em minus 0.4em\relax IEEE, 2016, pp. 1--6.

\bibitem{DroneDetectionSystem}
Aaronia, ``Drone detection system,'' \url{https://www.aaronia.com/}.

\bibitem{GROK}
Anti-Drone.eu, ``Grok,'' \url{https://anti-drone.eu/}.

\bibitem{Gamekeeper16U}
Aveillant, ``Gamekeeper 16u - holographic radar,''
  \url{http://www.aveillant.com/}.

\bibitem{UAVX}
B.~S. BST, ``Uavx,'' \url{https://www.blacksagetech.com/}.

\bibitem{LightWave}
C.~speed LLC, ``Lightwave radar,'' \url{http://cspeed.com/}.

\bibitem{AUDS}
C.~D. Ltd, ``Auds,'' \url{www.chess-dynamics.com/}.

\bibitem{HARRIER}
DeTect, ``Harrier dsr,'' \url{https://detect-inc.com/}.

\bibitem{Skylight}
G.~Sensors, ``Skylight,'' \url{https://www.srcinc.com/}.

\bibitem{SharpEve}
K.~H. Limited, ``Sharpeve sxv radar,'' \url{https://www.kelvinhughes.com/}.

\bibitem{ELVIRA}
R.~R. Systems, ``Elvira,'' \url{https://www.robinradar.com/}.

\bibitem{Giraffe}
S.~Group, ``Giraffe amb radar - elss,'' \url{https://saabgroup.com/}.

\bibitem{A2000}
SpotterRF, ``A2000 radar uavx,'' \url{https://spotterrf.com/}.

\bibitem{SQUIRE}
Thales, ``Squire,'' \url{www.thalesgroup.com/en}.

\bibitem{birnbach2017wi}
S.~Birnbach, R.~Baker, and I.~Martinovic, ``Wi-fly?: Detecting privacy invasion
  attacks by consumer drones,'' \emph{NDSS}, 2017.

\bibitem{nguyen2017matthan}
P.~Nguyen, H.~Truong, M.~Ravindranathan, A.~Nguyen, R.~Han, and T.~Vu,
  ``Matthan: Drone presence detection by identifying physical signatures in the
  drone's rf communication,'' in \emph{Proceedings of the 15th Annual
  International Conference on Mobile Systems, Applications, and
  Services}.\hskip 1em plus 0.5em minus 0.4em\relax ACM, 2017, pp. 211--224.

\bibitem{nguyen2018cost}
------, ``Cost-effective and passive rf-based drone presence detection and
  characterization,'' \emph{GetMobile: Mobile Computing and Communications},
  vol.~21, no.~4, pp. 30--34, 2018.

\bibitem{nguyen2016investigating}
P.~Nguyen, M.~Ravindranatha, A.~Nguyen, R.~Han, and T.~Vu, ``Investigating
  cost-effective rf-based detection of drones,'' in \emph{Proceedings of the
  2nd Workshop on Micro Aerial Vehicle Networks, Systems, and Applications for
  Civilian Use}.\hskip 1em plus 0.5em minus 0.4em\relax ACM, 2016, pp. 17--22.

\bibitem{scheller2017detecting}
W.~D. Scheller, ``Detecting drones using machine learning,'' 2017.

\bibitem{shi2017detection}
Z.~Shi, M.~Huang, C.~Zhao, L.~Huang, X.~Du, and Y.~Zhao, ``Detection of lssuav
  using hash fingerprint based svdd,'' in \emph{Communications (ICC), 2017 IEEE
  International Conference on}.\hskip 1em plus 0.5em minus 0.4em\relax IEEE,
  2017, pp. 1--5.

\bibitem{peacock2013towards}
M.~Peacock and M.~N. Johnstone, ``Towards detection and control of civilian
  unmanned aerial vehicles,'' 2013.

\bibitem{mototolea2018detection}
D.~Mototolea and C.~Stolk, ``Detection and localization of small drones using
  commercial off-the-shelf fpga based software defined radio systems,'' in
  \emph{2018 International Conference on Communications (COMM)}.\hskip 1em plus
  0.5em minus 0.4em\relax IEEE, 2018, pp. 465--470.

\bibitem{RogueDrone}
3DEO, ``Rogue drone detection and mitigation,''
  \url{https://3deo.biz/applications/drone-detection-and-mitigation}.

\bibitem{droneShield}
Anti-Drone.eu, ``droneshield,'' \url{https://anti-drone.eu/}.

\bibitem{SkyTracker}
CACI, ``Skytracker,'' \url{http://www.caci.com/}.

\bibitem{DroneWatch}
CerbAir, ``Dronewatch,'' \url{https://www.cerbair.com/}.

\bibitem{DroneTracker}
DeDrone.com, ``Dronetracker,'' \url{https://www.dedrone.com/}.

\bibitem{DroneWatcher}
DeTect, ``Dronewatcher,'' \url{https://detect-inc.com}.

\bibitem{SigBASE}
D.~G. Systems, ``Sigbase,'' \url{https://www.digitalglobalsystems.com/}.

\bibitem{FarAlert}
DroneShield, ``Faralert/widealet sensors,'' \url{https://www.droneshield.com/}.

\bibitem{UAVdetection}
H.~infrared systems, ``Uav detection and tracking,''
  \url{https://www.hgh-infrared.com/}.

\bibitem{DroneDetection2}
MAGNA, ``Drone detection,'' \url{https://magnabsp.com}.

\bibitem{Skysentry}
M.~AVISA, ``Skysentry amms,'' \url{microflown-avisa.com/}.

\bibitem{DroneDetection1}
M.~Solutions, ``Drone detection and classification system,''
  \url{https://www.mistralsolutions.com/}.

\bibitem{DroneDetector}
ORELIA, ``Drone-detector,'' \url{http://www.drone-detector.com/en/}.

\bibitem{Qguard}
Q.~Systems, ``Q-guard - lidar x-drone,'' \url{https://quanergy.com/}.

\bibitem{SkyPATRIOT}
Rinicom, ``Sky patriot,'' \url{www.rinicom.com/}.

\bibitem{SkyPERION}
Rinicom and M.~Aerospace, ``Skyperion,''
  \url{http://metisaerospace.com/skyperion-counter-uav/}.

\bibitem{ARDRONIS}
R.~. Schwarz, ``R\&s ardronis-i,'' \url{https://www.rohde-schwarz.com}.

\bibitem{AIRFENCE}
Sensofusion, ``Airfence,'' \url{https://www.sensofusion.com/}.

\bibitem{DiscovAir}
S.~Technology, ``Discovair,'' \url{www.sqhead.com/}.

\bibitem{BlackBird}
T.~International, ``Blackbird,'' \url{https://www.tcibr.com/}.

\bibitem{hu2017detection}
S.~Hu, G.~H. Goldman, and C.~C. Borel-Donohue, ``Detection of unmanned aerial
  vehicles using a visible camera system,'' \emph{Applied optics}, vol.~56,
  no.~3, pp. B214--B221, 2017.

\bibitem{rozantsev2015flying}
A.~Rozantsev, V.~Lepetit, and P.~Fua, ``Flying objects detection from a single
  moving camera,'' in \emph{Proceedings of the IEEE Conference on Computer
  Vision and Pattern Recognition}, 2015, pp. 4128--4136.

\bibitem{santana2014trajectory}
L.~V. Santana, A.~S. Brandao, M.~Sarcinelli-Filho, and R.~Carelli, ``A
  trajectory tracking and 3d positioning controller for the ar. drone
  quadrotor,'' in \emph{Unmanned Aircraft Systems (ICUAS), 2014 International
  Conference on}.\hskip 1em plus 0.5em minus 0.4em\relax IEEE, 2014, pp.
  756--767.

\bibitem{unlu2018using}
E.~Unlu, E.~Zenou, and N.~Rivi{\`e}re, ``Using shape descriptors for uav
  detection,'' \emph{Electronic Imaging}, vol. 2018, no.~9, pp. 1--5, 2018.

\bibitem{saqib2017study}
M.~Saqib, S.~D. Khan, N.~Sharma, and M.~Blumenstein, ``A study on detecting
  drones using deep convolutional neural networks,'' in \emph{2017 14th IEEE
  International Conference on Advanced Video and Signal Based Surveillance
  (AVSS)}.\hskip 1em plus 0.5em minus 0.4em\relax IEEE, 2017, pp. 1--5.

\bibitem{aker2017using}
C.~Aker and S.~Kalkan, ``Using deep networks for drone detection,'' \emph{arXiv
  preprint arXiv:1706.05726}, 2017.

\bibitem{unlu2018generic}
E.~Unlu, E.~Zenou, and N.~Rivi{\`e}re, ``Generic fourier descriptors for
  autonomous uav detection,'' 2018.

\bibitem{rozantsev2017flight}
A.~Rozantsev, S.~N. Sinha, D.~Dey, and P.~Fua, ``Flight dynamics-based recovery
  of a uav trajectory using ground cameras,'' in \emph{Conf. Comp. Vision and
  Pattern Recognition}, 2017.

\bibitem{muller2017robust}
T.~M{\"u}ller, ``Robust drone detection for day/night counter-uav with static
  vis and swir cameras,'' in \emph{Ground/Air Multisensor Interoperability,
  Integration, and Networking for Persistent ISR VIII}, vol. 10190.\hskip 1em
  plus 0.5em minus 0.4em\relax International Society for Optics and Photonics,
  2017, p. 1019018.

\bibitem{birch2017counter}
G.~C. Birch and B.~L. Woo, ``Counter unmanned aerial systems testing:
  Evaluation of vis swir mwir and lwir passive imagers,'' 2017.

\bibitem{church2018aerial}
P.~Church, C.~Grebe, J.~Matheson, and B.~Owens, ``Aerial and surface security
  applications using lidar,'' in \emph{Laser Radar Technology and Applications
  XXIII}, vol. 10636.\hskip 1em plus 0.5em minus 0.4em\relax International
  Society for Optics and Photonics, 2018, p. 1063604.

\bibitem{kim2017real}
J.~Kim, C.~Park, J.~Ahn, Y.~Ko, J.~Park, and J.~C. Gallagher, ``Real-time uav
  sound detection and analysis system,'' in \emph{Sensors Applications
  Symposium (SAS), 2017 IEEE}.\hskip 1em plus 0.5em minus 0.4em\relax IEEE,
  2017, pp. 1--5.

\bibitem{mezei2016drone}
J.~Mezei and A.~Moln{\'a}r, ``Drone sound detection by correlation,'' in
  \emph{Applied Computational Intelligence and Informatics (SACI), 2016 IEEE
  11th International Symposium on}.\hskip 1em plus 0.5em minus 0.4em\relax
  IEEE, 2016, pp. 509--518.

\bibitem{mirelli2009statistical}
V.~Mirelli, S.~Tenney, Y.~Bengio, N.~Chapados, and O.~Delalleau, ``Statistical
  machine learning algorithms for target classification from acoustic
  signature,'' in \emph{Proc. MSS Battlespace Acoust. Magn. Sensors}, 2009, pp.
  1--18.

\bibitem{roseveare2006robust}
N.~J. Roseveare and M.~R. Azimi-Sadjadi, ``Robust beamforming algorithms for
  acoustic tracking of ground vehicles,'' in \emph{Unattended Ground, Sea, and
  Air Sensor Technologies and Applications VIII}, vol. 6231.\hskip 1em plus
  0.5em minus 0.4em\relax International Society for Optics and Photonics, 2006,
  p. 623107.

\bibitem{chang2018surveillance}
X.~Chang, C.~Yang, J.~Wu, X.~Shi, and Z.~Shi, ``A surveillance system for drone
  localization and tracking using acoustic arrays,'' in \emph{2018 IEEE 10th
  Sensor Array and Multichannel Signal Processing Workshop (SAM)}.\hskip 1em
  plus 0.5em minus 0.4em\relax IEEE, 2018, pp. 573--577.

\bibitem{sedunov2016passive}
A.~Sedunov, A.~Sutin, N.~Sedunov, H.~Salloum, A.~Yakubovskiy, and D.~Masters,
  ``Passive acoustic system for tracking low-flying aircraft,'' \emph{IET
  Radar, Sonar \& Navigation}, vol.~10, no.~9, pp. 1561--1568, 2016.

\bibitem{yakubovskiy2015feature}
A.~Yakubovskiy, H.~Salloum, A.~Sutin, A.~Sedunov, N.~Sedunov, and D.~Masters,
  ``Feature extraction for acoustic classification of small aircraft,'' in
  \emph{Applications of Signal Processing to Audio and Acoustics (WASPAA), 2015
  IEEE Workshop on}.\hskip 1em plus 0.5em minus 0.4em\relax IEEE, 2015, pp.
  1--5.

\bibitem{case2008low}
E.~E. Case, A.~M. Zelnio, and B.~D. Rigling, ``Low-cost acoustic array for
  small uav detection and tracking,'' in \emph{Aerospace and Electronics
  Conference, 2008. NAECON 2008. IEEE National}.\hskip 1em plus 0.5em minus
  0.4em\relax IEEE, 2008, pp. 110--113.

\bibitem{hommes2016detection}
A.~Hommes, A.~Shoykhetbrod, D.~Noetel, S.~Stanko, M.~Laurenzis, S.~Hengy, and
  F.~Christnacher, ``Detection of acoustic, electro-optical and radar
  signatures of small unmanned aerial vehicles,'' in \emph{Target and
  Background Signatures II}, vol. 9997.\hskip 1em plus 0.5em minus 0.4em\relax
  International Society for Optics and Photonics, 2016, p. 999701.

\bibitem{laurenzis2017multi}
M.~Laurenzis, S.~Hengy, A.~Hommes, F.~Kloeppel, A.~Shoykhetbrod, T.~Geibig,
  W.~Johannes, P.~Naz, and F.~Christnacher, ``Multi-sensor field trials for
  detection and tracking of multiple small unmanned aerial vehicles flying at
  low altitude,'' in \emph{Signal Processing, Sensor/Information Fusion, and
  Target Recognition XXVI}, vol. 10200.\hskip 1em plus 0.5em minus 0.4em\relax
  International Society for Optics and Photonics, 2017, p. 102001A.

\bibitem{liu2017drone}
H.~Liu, Z.~Wei, Y.~Chen, J.~Pan, L.~Lin, and Y.~Ren, ``Drone detection based on
  an audio-assisted camera array,'' in \emph{Multimedia Big Data (BigMM), 2017
  IEEE Third International Conference on}.\hskip 1em plus 0.5em minus
  0.4em\relax IEEE, 2017, pp. 402--406.

\bibitem{busset2015detection}
J.~Busset, F.~Perrodin, P.~Wellig, B.~Ott, K.~Heutschi, T.~R{\"u}hl, and
  T.~Nussbaumer, ``Detection and tracking of drones using advanced acoustic
  cameras,'' in \emph{Unmanned/Unattended Sensors and Sensor Networks XI; and
  Advanced Free-Space Optical Communication Techniques and Applications}, vol.
  9647.\hskip 1em plus 0.5em minus 0.4em\relax International Society for Optics
  and Photonics, 2015, p. 96470F.

\bibitem{christnacher2016optical}
F.~Christnacher, S.~Hengy, M.~Laurenzis, A.~Matwyschuk, P.~Naz, S.~Schertzer,
  and G.~Schmitt, ``Optical and acoustical uav detection,'' in
  \emph{Electro-Optical Remote Sensing X}, vol. 9988.\hskip 1em plus 0.5em
  minus 0.4em\relax International Society for Optics and Photonics, 2016, p.
  99880B.

\bibitem{shi2011detecting}
W.~Shi, G.~Arabadjis, B.~Bishop, P.~Hill, R.~Plasse, and J.~Yoder, ``Detecting,
  tracking, and identifying airborne threats with netted sensor fence,'' in
  \emph{Sensor Fusion-Foundation and Applications}.\hskip 1em plus 0.5em minus
  0.4em\relax InTech, 2011.

\bibitem{vasquez2008multisensor}
J.~R. Vasquez, K.~M. Tarplee, E.~E. Case, A.~M. Zelnio, and B.~D. Rigling,
  ``Multisensor 3d tracking for counter small unmanned air vehicles (csuav),''
  in \emph{Acquisition, Tracking, Pointing, and Laser Systems Technologies
  XXII}, vol. 6971.\hskip 1em plus 0.5em minus 0.4em\relax International
  Society for Optics and Photonics, 2008, p. 697107.

\bibitem{boddhu2013collaborative}
S.~K. Boddhu, M.~McCartney, O.~Ceccopieri, and R.~L. Williams, ``A
  collaborative smartphone sensing platform for detecting and tracking hostile
  drones,'' \emph{SPIE Defense, Security, and Sensing}, pp. 874\,211--874\,211,
  2013.

\bibitem{irondome-5-million}
Forbes, ``British army used israeli tech to end gatwick airport xmas drone
  chaos,''
  \url{https://www.forbes.com/sites/annatobin/2018/12/26/british-army-used-israeli-tech-to-end-gatwick-airport-xmas-drone-chaos/#75b0793f6e6e}.

\bibitem{nassi2018game}
B.~Nassi, R.~Ben-Netanel, A.~Shamir, and Y.~Elovici, ``Game of drones-detecting
  streamed poi from encrypted fpv channel,'' \emph{arXiv preprint
  arXiv:1801.03074}, 2018.

\bibitem{nassi-smashing}
\BIBentryALTinterwordspacing
------, ``Drones' cryptanalysis - smashing cryptography with a flicker,'' in
  \emph{2019 2019 IEEE Symposium on Security and Privacy (SP)}.\hskip 1em plus
  0.5em minus 0.4em\relax Los Alamitos, CA, USA: IEEE Computer Society, may
  2019. [Online]. Available:
  \url{https://doi.ieeecomputersociety.org/10.1109/SP.2019.00051}
\BIBentrySTDinterwordspacing

\bibitem{dronescrypta}
\BIBentryALTinterwordspacing
------, ``Drones' cryptanalysis - smashing cryptography with a flicker,'' in
  \emph{2019 2019 IEEE Symposium on Security and Privacy (SP)}, vol.~00, pp.
  833--850. [Online]. Available:
  \url{doi.ieeecomputersociety.org/10.1109/SP.2019.00051}
\BIBentrySTDinterwordspacing

\bibitem{Ruiz:2018:IRD:3236498.3214283}
\BIBentryALTinterwordspacing
C.~Ruiz, S.~Pan, A.~Bannis, X.~Chen, C.~Joe-Wong, H.~Y. Noh, and P.~Zhang,
  ``Idrone: Robust drone identification through motion actuation feedback,''
  \emph{Proc. ACM Interact. Mob. Wearable Ubiquitous Technol.}, vol.~2, no.~2,
  pp. 80:1--80:22, Jul. 2018. [Online]. Available:
  \url{http://doi.acm.org/10.1145/3214283}
\BIBentrySTDinterwordspacing

\bibitem{friend-or-foe}
Wikipedia, ``Identification friend or foe,''
  \url{https://en.wikipedia.org/wiki/Identification_friend_or_foe}.

\bibitem{GPS-Accuracy}
``Gps accuracy,'' \url{https://www.gps.gov/systems/gps/performance/accuracy/}.

\bibitem{SkyJack}
S.~Kamkar, ``Skyjack,'' 2015.

\bibitem{Knocking-my-neighbors-kids-cruddy-drone-offline}
M.~Robinson, ``Knocking my neighbors kids cruddy drone offline,'' \emph{DEF CON
  23}, 2016.

\bibitem{Deligne2012}
\BIBentryALTinterwordspacing
E.~Deligne, ``Ardrone corruption,'' \emph{Journal in Computer Virology},
  vol.~8, no.~1, pp. 15--27, May 2012. [Online]. Available:
  \url{https://doi.org/10.1007/s11416-011-0158-4}
\BIBentrySTDinterwordspacing

\bibitem{Hacking-A-Professional-Drone}
N.~Rodday, ``Hacking a professional drone,'' \emph{Black Hat Asia}, 2016.

\bibitem{highnam2016uncrewed}
K.~Highnam, K.~Angstadt, K.~Leach, W.~Weimer, A.~Paulos, and P.~Hurley, ``An
  uncrewed aerial vehicle attack scenario and trustworthy repair
  architecture,'' in \emph{2016 46th Annual IEEE/IFIP International Conference
  on Dependable Systems and Networks Workshop (DSN-W)}.\hskip 1em plus 0.5em
  minus 0.4em\relax IEEE, 2016, pp. 222--225.

\bibitem{shoufan2017continuous}
A.~Shoufan, ``Continuous authentication of uav flight command data using
  behaviometrics,'' in \emph{2017 IFIP/IEEE International Conference on Very
  Large Scale Integration (VLSI-SoC)}.\hskip 1em plus 0.5em minus 0.4em\relax
  IEEE, 2017, pp. 1--6.

\bibitem{davidson2016controlling}
D.~Davidson, H.~Wu, R.~Jellinek, V.~Singh, and T.~Ristenpart, ``Controlling
  uavs with sensor input spoofing attacks.'' in \emph{WOOT}, 2016.

\bibitem{Drones-Hijacking}
A.~Luo, ``Drones hijacking - multi-dimensional attack vectors and
  countermeasures,'' \emph{DefCon 24}.

\bibitem{kerns2014unmanned}
A.~J. Kerns, D.~P. Shepard, J.~A. Bhatti, and T.~E. Humphreys, ``Unmanned
  aircraft capture and control via gps spoofing,'' \emph{Journal of Field
  Robotics}, vol.~31, no.~4, pp. 617--636, 2014.

\bibitem{feng2017efficient}
Z.~Feng, N.~Guan, M.~Lv, W.~Liu, Q.~Deng, X.~Liu, and W.~Yi, ``Efficient drone
  hijacking detection using onboard motion sensors,'' in \emph{Design,
  Automation \& Test in Europe Conference \& Exhibition (DATE), 2017}.\hskip
  1em plus 0.5em minus 0.4em\relax IEEE, 2017, pp. 1414--1419.

\bibitem{feng2018efficient}
------, ``An efficient uav hijacking detection method using onboard inertial
  measurement unit,'' \emph{ACM Transactions on Embedded Computing Systems
  (TECS)}, vol.~17, no.~6, p.~96, 2018.

\bibitem{son2015rocking}
Y.~Son, H.~Shin, D.~Kim, Y.-S. Park, J.~Noh, K.~Choi, J.~Choi, Y.~Kim
  \emph{et~al.}, ``Rocking drones with intentional sound noise on gyroscopic
  sensors.'' in \emph{USENIX Security Symposium}, 2015, pp. 881--896.

\bibitem{choi2018detecting}
H.~Choi, W.-C. Lee, Y.~Aafer, F.~Fei, Z.~Tu, X.~Zhang, D.~Xu, and X.~Xinyan,
  ``Detecting attacks against robotic vehicles: A control invariant approach,''
  in \emph{Proceedings of the 2018 ACM SIGSAC Conference on Computer and
  Communications Security}.\hskip 1em plus 0.5em minus 0.4em\relax ACM, 2018,
  pp. 801--816.

\bibitem{206156}
\BIBentryALTinterwordspacing
S.~Belikovetsky, M.~Yampolskiy, J.~Toh, J.~Gatlin, and Y.~Elovici, ``dr0wned
  {\textendash} cyber-physical attack with additive manufacturing,'' in
  \emph{11th {USENIX} Workshop on Offensive Technologies ({WOOT} 17)}.\hskip
  1em plus 0.5em minus 0.4em\relax Vancouver, BC: {USENIX} Association, 2017.
  [Online]. Available:
  \url{https://www.usenix.org/conference/woot17/workshop-program/presentation/belikovetsky}
\BIBentrySTDinterwordspacing

\bibitem{parazero}
ParaZero, ``Safeairtm m-200,''
  \url{https://parazero.com/solutions/safeair-for-dji-matrice-200/}.

\bibitem{MARS-Parachutes}
M.~Parachutes, ``Mars parachutes,'' \url{https://www.marsparachutes.com/}.

\bibitem{fruitychutes}
F.~Chutes, ``Ultimate parachute system for all drones, multicopters and uas,''
  \url{https://fruitychutes.com/uav_rpv_drone_recovery_parachutes.htm}.

\bibitem{Maldrone}
R.~Sasi, ``Maldrone,'' \url{http://garage4hackers.com/entry.php?b=3105}, 2015.

\bibitem{pike2016trackos}
L.~Pike, P.~Hickey, T.~Elliott, E.~Mertens, and A.~Tomb, ``Trackos: A
  security-aware real-time operating system,'' in \emph{International
  Conference on Runtime Verification}.\hskip 1em plus 0.5em minus 0.4em\relax
  Springer, 2016, pp. 302--317.

\bibitem{trippel2017walnut}
T.~Trippel, O.~Weisse, W.~Xu, P.~Honeyman, and K.~Fu, ``Walnut: Waging doubt on
  the integrity of mems accelerometers with acoustic injection attacks,'' in
  \emph{Security and Privacy (EuroS\&P), 2017 IEEE European Symposium
  on}.\hskip 1em plus 0.5em minus 0.4em\relax IEEE, 2017, pp. 3--18.

\bibitem{farshteindiker2016phone}
B.~Farshteindiker, N.~Hasidim, A.~Grosz, and Y.~Oren, ``How to phone home with
  someone else's phone: Information exfiltration using intentional sound noise
  on gyroscopic sensors.'' in \emph{WOOT}, 2016.

\bibitem{vervisch2017influence}
A.~Vervisch-Picois, N.~Samama, and T.~Taillandier-Loize, ``Influence of gnss
  spoofing on drone in automatic flight mode,'' in \emph{ITSNT 2017: 4th
  International Symposium of Navigation and Timing}.\hskip 1em plus 0.5em minus
  0.4em\relax Ecole nationale de l'aviation civile, 2017, pp. 1--9.

\bibitem{he2018friendly}
D.~He, Y.~Qiao, S.~Chen, X.~Du, W.~Chen, S.~Zhu, and M.~Guizani, ``A friendly
  and low-cost technique for capturing non-cooperative civilian unmanned aerial
  vehicles,'' \emph{IEEE Network}, 2018.

\bibitem{DroneDefender}
Battelle, ``Dronedefender,''
  \url{https://www.battelle.org/government-offerings/national-security/aerospace-systems/counter-UAS-technologies/dronedefender}.

\bibitem{AUDS1}
Bilghter, ``Auds,''
  \url{http://www.blighter.com/products/auds-anti-uav-defence-system.html}.

\bibitem{DroneJammerGun}
C.~Technology, ``Drone jammer gun,''
  \url{https://ctstechnologys.com/c-t-s-developed-a-long-distance-drone-jammer.html}.

\bibitem{DroneGun}
DroneShield, ``Dronegun,''
  \url{https://www.droneshield.com/dronegun-tactical/}.

\bibitem{SKYNET}
H.~. MiGHTY, ``Skynet rifle,'' \url{http://anti-drones.net/}.

\bibitem{Scrambler}
M.~Technologies, ``Scrambler 300 rifle,''
  \url{https://defense-update.com/20170301_scrambler300.html}.

\bibitem{Dronebuster}
R.~Hill, ``Block 3 dronebuster and dronebuster-le,''
  \url{http://radiohill.com/2016/11/10/block-3-dronebuster-technologies/}.

\bibitem{Dynopis}
D.~D. UK, ``Dynopis r1000mp,''
  \url{https://www.dronedefence.co.uk/products/dynopis-e1000mp/}.

\bibitem{Xpeller}
Hensoldt, ``Xpeller c-uav,''
  \url{https://www.hensoldt.net/solutions/air/electronic-warfare/xpeller-counter-uav-system/}.

\bibitem{MCHORIZON}
MCTECH, ``Mc-horizon,''
  \url{http://mctech-jammers.com/products/mc-horizon.html}.

\bibitem{DroneDefeater}
S.~Group, ``Drone defeater,'' \url{https://www.sespgroup.com/drone-defeater/}.

\bibitem{OneTouchInterceptor}
A.~S. Inc, ``Drone detection system,'' \url{https://airspace.co/}.

\bibitem{DroneCatcher}
------, ``Delft dynamics,'' \url{https://www.delftdynamics.nl/}.

\bibitem{DroneHunter}
Fortem, ``Dronehunter,'' \url{https://fortemtech.com/products/dronehunter/}.

\bibitem{RoboticFalconry}
M.~Tech, ``Robotic falconry,'' \url{http://me.sites.mtu.edu/rastgaar/hirolab/}.

\bibitem{SparrowHawk}
S.~Systems, ``Sparrowhawk phase one u-uav,''
  \url{http://searchsystems.eu/sparrowhawk.html}.

\bibitem{NetGun}
C.~Enterprises, ``Net gun,''
  \url{http://www.codaenterprises.com/products.html}.

\bibitem{NetGunX1}
D.~D. UK, ``Net gun x1,''
  \url{https://www.dronedefence.co.uk/products/netgun-x1/}.

\bibitem{SkyWall}
O.~Engineering, ``Skywall 100 launcher,''
  \url{https://openworksengineering.com/skywall}.

\bibitem{ShotgunShellNets}
SkyNet, ``12 gauge shotgun shell nets for drones,''
  \url{https://www.budk.com/}.

\bibitem{DroneHuntingEagles}
G.~F. Above, ``Drone hunting eagles,'' \url{www.guardfromabove.com}.

\bibitem{IronBeam}
Rafael, ``Ironbeam,'' \url{http://www.rafael.co.il/5688-763-en/Marketing.aspx}.

\bibitem{LaserEffector}
MBDA, ``Laser effector,''
  \url{https://www.mbda-systems.com/innovation/preparing-future-products-3/high-energy-laser-weapon-systems/}.

\bibitem{patriot}
T.~Verge, ``A us ally shot down a \$200 drone with a \$3 million patriot
  missile,''
  \url{https://www.theverge.com/2017/3/16/14944256/patriot-missile-shot-down-consumer-drone-us-military},
  2017.

\bibitem{zeng2016wiwho}
Y.~Zeng, P.~H. Pathak, and P.~Mohapatra, ``Wiwho: wifi-based person
  identification in smart spaces,'' in \emph{Proceedings of the 15th
  International Conference on Information Processing in Sensor Networks}.\hskip
  1em plus 0.5em minus 0.4em\relax IEEE Press, 2016, p.~4.

\end{thebibliography}
\end{document}